\documentclass[aps,twocolumn,pra]{revtex4-2}
\usepackage[T1]{fontenc}
\usepackage{dsfont}
\usepackage{amsmath}
\usepackage{bm}
\usepackage{amssymb}
\usepackage{graphicx}
\usepackage{braket}
\usepackage{amsthm}
\usepackage[colorlinks=true,urlcolor=blue,citecolor=blue,linkcolor=blue]{hyperref}
\DeclareGraphicsExtensions{.eps,.ps,.pdf}
\newtheorem{wit}{Witness}
\newtheorem*{witghost}{Witness~\cite{Boh.Gir.Bra:17}}
\newcommand{\diff}{\mathop{}\!\mathrm{d}}
\newcommand{\al}{\alpha}
\newcommand{\la}{\lambda}
\newcommand{\si}{\sigma}
\newcommand{\La}{\Lambda}
\def \Tr {\text{Tr}}
\newcommand{\ie}{\hbox{\em i.e.{}}}

\newcommand{\rhs}{\hbox{r.h.s.{}}}
\newcommand{\Rr}{\mathsf{R}}
\newcommand{\Hs}{\mathcal{H}}
\newcommand{\BHs}{\mathcal{B}}
\newcommand{\da}{\dagger}

\newcommand{\trun}{0}
\newcommand{\LB}{LB}
\newcommand{\SAS}{\mathcal{A}_{\mathrm{sym}}}

\newcommand{\BIS}{\mathcal{B}}
\newcommand{\UNIS}{\mathcal{U}}
\newcommand{\WIT}{\mathcal{W}}
\newcommand{\dist}{r}
\newcommand{\Sep}{\mathcal{S}}
\newcommand{\Nonli}{\Pi}

\newcommand{\rema}[1]{\textcolor{black}{#1}}

\makeatletter
\g@addto@macro\bfseries{\boldmath}
\makeatother
\begin{document} 

\widetext

\title{Absolute separability witnesses for symmetric multiqubit states} 

\author{Eduardo Serrano-Ens\'astiga}
\email{ed.ensastiga@uliege.be}
\affiliation{Institut de Physique Nucléaire, Atomique et de Spectroscopie, CESAM, University of Liège
\\
B-4000 Liège, Belgium}

\author{Jérôme Denis}
\email{jdenis@uliege.be}
\affiliation{Institut de Physique Nucléaire, Atomique et de Spectroscopie, CESAM, University of Liège
\\
B-4000 Liège, Belgium}

\author{John Martin}
\email{jmartin@uliege.be}
\affiliation{Institut de Physique Nucléaire, Atomique et de Spectroscopie, CESAM, University of Liège
\\
B-4000 Liège, Belgium}

\date{January 19, 2024}
\begin{abstract}
The persistent separability of certain quantum states, known as \textit{symmetric absolutely separable} (SAS), under symmetry-preserving global unitary transformations is of key significance in the context of quantum resources for bosonic systems.
In this work, we develop criteria for detecting SAS states of any number of qubits. Our approach is based on the Glauber-Sudarshan $P$ representation for finite-dimensional quantum systems. We introduce three families of SAS witnesses, one linear and two nonlinear in the \rema{eigenvalues of the} state, formulated respectively as an algebraic inequality or a quadratic optimization problem. These witnesses are capable of identifying more SAS states than previously known counterparts. We also explore the geometric properties of the subsets of SAS states detected by our witnesses, shedding light on their distinctions.
\end{abstract}
\maketitle
\section{Introduction}
Quantum entanglement is a pivotal concept in the foundations of quantum theory, with significant implications for various quantum technology applications, including quantum cryptography, quantum sensing, metrology, and quantum simulation~\cite{DeutschJ92,Bennett93,Grover97,shor1999polynomial,
nielsen_chuang_2010,Ben.Zyc:17}. Understanding how to generate, detect and quantify entanglement is crucial, akin to managing any valuable resource. One established method for creating entanglement is to apply a global unitary transformation to a system of interest via a unitary channel or a smooth unitary evolution driven by a control Hamiltonian. Nonetheless, it is possible that even after such transformations, an initially separable state remains separable~\cite{PhysRevA.40.4277}, i.e.\ unentangled. The characterization of these states, known as absolutely separable (AS)~\cite{Kus.Zyc:01}, helps identify necessary conditions in a state's spectrum to turn it into an entangled state after a unitary gate. In particular, such characterizations can provide valuable insights about entangled states very close to the maximally mixed state~\cite{zyczkowski_volume_1998,Gur:02,PhysRevA.75.062330,Adh:21}. This question is particularly relevant in systems subjected to noisy environments~\cite{Hal:13,PhysRevA.88.062328} as seen, e.g., in NMR settings~\cite{Bra.Cav:99,JONES201191}. Determining whether the quantum state of a multipartite system is separable is a NP-\rema{hard} problem~\cite{10.1145/780542.780545,GUHNE20091}. Consequently, the absolute version of the problem may share the same complexity. So far, the characterization of AS states has been limited to qubit-qudit systems~\cite{verstraete_maximally_2001,PhysRevA.76.052325,PhysRevA.88.062330}. For higher multiqudit systems, there are several works defining witnesses of (not-)absolute separability~\cite{PhysRevA.104.032427,thirring:11,entropy.Leggio}. 

In addition to the problem of implementing unitary operations, the permissible unitary transformations on a quantum system are sometimes limited by its very physical nature, such as the physical properties of its constituents. In this work, we focus on permutation-invariant mixed states of a system composed of many qubits, i.e.\ symmetric multiqubit states. This type of states appears naturally in bosonic systems such as multiphotonic~\cite{Kok.Mun.Nem.Ral:07,Pan.Che:12} or atomic systems with a collective angular momentum, be it spin, orbital, or a combination of both. 
Formally speaking, the set of physical states valid for bosonic systems reduces de facto to the symmetric subspace $\Hs_1^{\vee N}$ of the full Hilbert space $\Hs_1^{\otimes^N }$ formed by the tensor product of $N$ single-qubit spaces $\Hs_1$.
This symmetric subspace has dimension $N+1$, indicating that the global unitary transformations permitted are limited to operations of the group $SU(N+1)$, acting solely within $\Hs_1^{\vee N}$, instead of the whole $SU(2^N)$ unitary group. \rema{The entanglement in bosonic systems has been shown to be also a relevant resource for applications in quantum metrology and quantum information~\cite{PhysRevX.10.041012}}. 

In the context described above, the question of absolute separability arises in the following form: which symmetric states do maintain their separability even after undergoing any global symmetry-preserving unitary transformation? These particular states are referred to as Symmetric Absolutely Separable (SAS) states~\cite{cha.joh.mac:21,ser.mar:23}, or Absolutely Classical states in the framework of spin systems~\cite{Boh.Gir.Bra:17} (as elaborated in Section~\ref{concepts.Sec}). Previous research has provided a complete characterization of SAS states for two-qubit systems, and a numerical exploration has been conducted for three-qubit systems~\cite{ser.mar:23}. For an arbitrary number of qubits $N$, a nonlinear SAS witness based on a purity condition of the state has been derived in a prior study~\cite{Boh.Gir.Bra:17}. The primary objective of the present work is to introduce novel SAS witnesses based on 
the Glauber-Sudarshan representation for spin states~\cite{PhysRevA.6.2211,Gir.Bra.Bra:08}. This representation, originally developed for the quantum harmonic oscillator~\cite{PhysRevLett.10.277,PhysRev.131.2766}, provides a $P$ representation of symmetric multiqubit states useful to describe quantum states in terms of classical-like probability distributions and study the separability problem~\cite{Gir.Bra.Bra:08,Boh.Gir.Bra:17}. 

The absolute symmetric version of properties other than separability has also been explored, such as absolutely positive Wigner functions states~\cite{abbasli2020measures,abgaryan2020global,denis.martin:23} or absolutely symmetric PPT (Positive Partial Transpose) states when the partial transposed state is also restricted to the symmetric subspace~\cite{cha.joh.mac:21}. \rema{Additionally, it has been shown that an arbitrary qubit-qudit state is absolutely PPT if and only if it is AS~\cite{PhysRevA.88.062330}. For a general system, however, it is only known that absolute separability implies absolutely PPT, a natural consequence of the Peres criterion~\cite{PhysRevLett.77.1413}. }

This work is organized as follows: Section~\ref{concepts.Sec} reviews the necessary physical concepts and mathematical methods. In sections~\ref{Sec.2.W1} and \ref{Sec.3.W2}, we calculate and formulate three novel families of SAS witnesses. We explain the methods to calculate the witness functionals for two- and three-qubit cases in Sec.~\ref{Sec.5}. We analyze the sets of SAS states detected by these witnesses for a general number of qubits and compare them with each other in Sec.~\ref{Sec.4.disc}. Finally, we conclude with some closing remarks in Sec.~\ref{Sec.5.Con}.
\begin{table*}[t!]
    \centering
    \begin{tabular}{c@{\hskip 0.5in} c}
      Spin-$j$ states  & Symmetric $N$-qubit states 
      \\
      \hline
      \hline      
       Spin quantum number $j=\frac{N}{2}$  & Number of qubits $N=2j$
       \\[0.1cm]
       Standard $J_z$-eigenbasis $ \ket{j,m} \leftrightarrow \ket{D^{(j-m)}_{2j}}$  &
       Symmetric Dicke states $\ket{D^{(k)}_N} \leftrightarrow \ket{ \frac{N}{2} , \frac{N}{2}-k }$
       \\[0.1cm]
       Spin-coherent states $ \ket{\Omega} = D(\Omega) \ket{j,j}$ & 
       Symmetric product states $\ket{\Omega} = D(\Omega) \ket{D^{(0)}_N} = D(\Omega) \ket{+}^{\otimes N} $
    \end{tabular}
    \caption{Some relations on the correspondence between  the spaces of spin-$j$ states $\Hs^{(j)}$ and symmetric $N$-qubit states $\Hs_1^{\vee N}$.}
    \label{tab.Spin.Product}
\end{table*}
\section{Concepts and methods}
\label{concepts.Sec}
\subsection{Equivalence between symmetric multiqubit states and spin-\texorpdfstring{$j$}{Lg} states}
\label{Sub.spin.multi}
Starting with the Hilbert space $\Hs_1$ of a single qubit system, which is spanned by two basis vectors, denoted as $\ket{+}$ and $\ket{-}$, we proceed to define the Hilbert space for a collection of $N$ qubits, $\Hs_N=\Hs_1^{\otimes N}$, spanned by the $2^N$ product states 
$\ket{s_1} \otimes \dots \otimes \ket{s_N}$ with $s_k \in\{+,-\}$ for $k=1,\ldots, N$. For our purposes, we will consider only the symmetric subspace $\Hs_1^{\vee N}$, of dimension $N+1$, spanned by the symmetric Dicke states $\ket{D_{N}^{(k)}}$ defined as \cite{PhysRev.93.99}
\begin{equation}
    \ket{D_{N}^{(k)}} = K \sum_{\pi \in S_N} \pi \big( \underbrace{ \ket{+} \otimes \cdots \ket{+} }_{N-k} \otimes \underbrace{\ket{-} \otimes \cdots \ket{-} }_{k} \big)
\end{equation}
for $k=0,\ldots,N$, where $K$ is a normalization constant and the sum runs over all permutations of $N$ elements $S_N$. 
In $\Hs_1^{\vee N}$, the product states have the form $\otimes^N \ket{\psi}$, with $\ket{\psi} = U\ket{+}$ and $U\in SU(2)$. They are generally referred to as spin coherent states (SC)~\cite{Rad:71}, due to the isomorphism between the symmetric $N$-qubit space $\Hs_1^{\vee N}$ and the spin-$j=N/2 $ Hilbert space $\Hs^{(j)}$ (see Table~\ref{tab.Spin.Product}). The rule of correspondence associates the basis vectors of $\Hs_1$ with the $J_z$ eigenbasis of a spin-1/2 system, $\ket{\pm} = \ket{1/2 ,\pm 1/2}$. Consequently, the Dicke basis corresponds to the eigenbasis of the $J_z$ angular momentum operator in the $(2j+1)$-dimensional irreducible representation (spin-$j$ irrep), $\ket{D_N^{(j-m)}} = \ket{j,m}$. The diagonal $SU(2)$ transformations on the symmetric multiqubit states are translated to rigid rotations according to the spin-$j$ irrep $D^{(j)}(\Rr)$, or just $D(\Rr)$ for short. \rema{The entries of these $(N+1)\times (N+1)$} matrices can be expressed in the Euler angles as \rema{$D^{(j)}_{m' m}(\alpha, \, \beta , \, \gamma) = \bra{j,m'} e^{-i \al J_z} e^{-i \beta J_y} e^{-i \gamma J_z} \ket{j,m}$}~\cite{Var.Mos.Khe:88}. Another correspondence given by the isomorphism is between the symmetric product states and the spin-coherent states $D(\Rr)\ket{+}^{\otimes N} = D(\Rr) \ket{j,j}$. The symmetric product states can then be parametrized just by the orientation $\Omega$ of the rotated quantization axis
\begin{equation}
\label{Sta.sep.sym}
\ket{\Omega} \equiv
 D^{(j)}(\Omega) \ket{j,j} =  D^{(j)}(\Omega) \ket{+}^{\otimes N}  \, ,
\end{equation}
with $D^{(j)}(\Omega)= D^{(j)}(\phi , \, \theta , \, 0)$ the rotation that reorients the $z$ axis to the $n$ axis with spherical angles $\Omega= (\theta, \phi)$~\cite{Var.Mos.Khe:88}. We can conclude by the previous observation that the symmetric product states constitute a two-dimensional sphere $S^2$. We summarise the correspondence between symmetric multiqubit states and spin-$j$ states in Table~\ref{tab.Spin.Product}.  While most of the discussion throughout the text is explained in the multiqubit framework, the calculations are mostly done in the spin language where the $SU(2)$-irreps appear naturally.
\subsection{Multipole operators}
We summarise below the basic properties of the set of multipole operators $\{T_{LM} \}$, with $L=0, \dots, 2j$ and $M=-L, \dots ,L$~\cite{Fan:53,Var.Mos.Khe:88}. They form an orthonormal basis in the space of operators acting on spin-$j$ states, $\BHs (\Hs^{(j)})$, \ie
\begin{equation}
\begin{aligned}
& \Tr ( T_{L_1 M_1}^{\dagger} T_{L_2 M_2} ) = \delta_{L_1 L_2} \delta_{M_1 M_2}  \, .
\end{aligned}
\end{equation}
In addition, they satisfy
\begin{equation}
T_{L M}^{\dagger} = (-1)^{M} T_{L \,-M} \, 
\end{equation}
and transform block-diagonally under rotations $\Rr \in SO(3)$ according to the spin-$L$ irrep
 $D^{(L)}(\Rr)$,
\begin{equation}
\label{prop.sh}
D^{(j)}(\Rr)\, T_{L M}\, D^{(j) \dagger}(\Rr)  = \sum\limits_{M'=-L}^{L} D_{ M' M}^{(L)}(\Rr)\, T_{L M'} \, .
\end{equation}
The $T_{LM}$ operators can be given explicitly in terms of the Clebsch-Gordan
coefficients $C_{j_1 m_1 j_2 m_2}^{j m}$ as follows~\cite{Var.Mos.Khe:88}
\begin{equation}
\label{decomp.TensOp}
T_{L M}
=
\sum_{m,m'=-j}^j (-1)^{j-m'} C_{jm,j-m'}^{L M}\ket{j,m}\bra{j,m'} 
\, .
\end{equation}
A mixed spin-$j$ state $\rho$ can always be expanded in the $T_{LM}$ basis, which gives
\begin{equation}\label{rhoTLMexpansion}
    \rho = \sum_{L=0}^{2j} \sum_{M=-L}^L \rho_{LM} T_{LM} = \rho_0 + \sum_{L=1}^{2j} \sum_{M=-L}^L \rho_{LM} T_{LM}  \, , 
\end{equation}
with $\rho_{LM} = \Tr (\rho T^{\dagger}_{LM})$ and $\rho_0 = (2j+1)^{-1} \mathds{1}$ the maximally mixed state in the symmetric sector. The hermiticity of the density matrix enforces that \rema{$\rho^{*}_{LM} = (-1)^M \rho_{L -M} $}. We will denote by $r$ the Hilbert-Schmidt distance between $\rho$ and $\rho_0$. It reads
\begin{equation}
\label{distance.rho.rho0}
\begin{aligned}
    r \equiv \big\lVert \rho - \rho_0 \big\rVert_{\text{HS}} = & \, \sqrt{\Tr(\rho^2) - (2j+1)^{-1}} 
\end{aligned}
\end{equation}
or, in terms of $\rho_{LM}$, as
\begin{equation}
\label{Purity.cond}
r = \sqrt{\sum_{L=1}^{2j} \left(
\rho_{L0}^2 + 2\sum_{M=1}^L |\rho_{LM} |^2
\right)}
\, .
\end{equation}
\subsection{Separability in terms of \texorpdfstring{$P$}{Lg} representation}
The Glauber-Sudarshan representation for spin states involves expressing the density operator in a form that resembles a classical probability distribution function on the sphere~\cite{PhysRevA.6.2211}
\begin{equation}
\label{rho.P.one}
    \rho = \int_{S^2} P(\rho , \Omega) \ket{\Omega} \bra{\Omega} \diff \Omega
\end{equation}
where $\diff \Omega = \diff \mu(\Omega)$ is the area element of the sphere $S^2$, and with $\ket{\Omega}$ given by Eq.~\eqref{Sta.sep.sym}. The expression \eqref{rho.P.one} is called a \emph{$P$ representation of $\rho$}. The correspondence described in Subsection~\ref{Sub.spin.multi} allows us to establish a $P$ representation of symmetric multiqubit states $\rho \in \BHs (\Hs_1^{\vee N})$ straightforwardly
\begin{equation}
\label{rho.P.one.multiqubit}
    \rho = \int_{S^2} P(\rho , \Omega) \big(D(\Omega)\ket{+} \bra{+} D^\dagger(\Omega)\big)^{\otimes N} \diff \Omega \, .
\end{equation}
We are interested in \emph{separable} multiqubit states~\cite{PhysRevA.40.4277}, i.e.\ those that can be written in terms of a classical (\emph{positive}) probability distribution \rema{  over the whole set of product states $D(\Omega_1) \ket{+} \otimes \dots \otimes  D(\Omega_N) \ket{+}\in\Hs_1^{\otimes N}$. However, we do not need to consider the whole set for symmetric states. Indeed, it was shown in Refs.~\cite{PhysRevLett.95.120502,PhysRevA.94.042343} that a symmetric state is separable if and only if it can be written in terms of a positive probability distribution over the smaller set of symmetric product states $(D(\Omega) \ket{+} )^{\otimes N }$.} It follows that a symmetric multiqubit state $\rho$ is separable if and only if there exists a $P$ representation of the form~\eqref{rho.P.one.multiqubit} with a positive $P$ function defined on the sphere 
\begin{equation}
\label{positivityP}
P(\rho, \Omega) \geqslant 0\qquad \forall\;\Omega\in S^2.
\end{equation}
In the language of spin states, the question of separability is that of classicality because the symmetric product states correspond to spin-coherent states (see Table~\ref{tab.Spin.Product}), considered to be the most classical spin states~\cite{Rad:71,Gir.Bra.Bra:08}.

The most general $P$ representation of a state $\rho \in \BHs (\Hs_1^{\vee N})$ has the form~\cite{Gir.Bra.Bra:08} 
\begin{equation}
\label{general.Pfunction}
P(\rho , \Omega) = P_{\trun} (\rho , \, \Omega) + \sum_{L=2j+1}^{+\infty} \sum_{M=-L}^L x_{LM} Y_{LM}(\Omega) \, ,
\end{equation}
where $P_{\trun} (\rho , \, \Omega)$ is uniquely defined for $\rho$ and given a posteriori by Eqs.~\eqref{Trun.P0} and \eqref{Q.function}, $x_{LM}$ are complex numbers, and $Y_{LM}(\Omega)$ are spherical harmonics. The $P$ function of a state is not unique because the variables $x_{LM} $ can be any complex number provided that \eqref{general.Pfunction} is real and covariant under rotations~\cite{klimov2017gen},
\begin{equation}
\label{covariance.P}
P(\rho,  \Omega) = P(D(\Omega)^{\dagger} \rho D(\Omega), 0) .
\end{equation}
On the other hand, $P_{\trun} (\rho , \, \Omega)$ is a function uniquely defined for each $\rho$, referred here as the \emph{truncated $P$ function} of $\rho$ and given by
\begin{equation}
\label{Trun.P0}
    P_{\trun} (\rho, \Omega)
    = \sqrt{\frac{4\pi}{2j+1}} \sum_{L=0}^{2j} \left( C_{j j L0}^{j j} \right)^{-1} Q_L(\rho,\Omega) \, ,
\end{equation}
with real functions (by the hermiticity of $\rho$)
\begin{equation}
\label{Q.function}
Q_L(\rho , \Omega) = \sum_{M=-L}^L \rho_{LM} Y_{LM} (\Omega)
\end{equation}
which are covariant under rotations, just as $P$ is (see Eq.~\eqref{covariance.P}). The latter equations come from the expansion of the operator $\ket{\Omega}\bra{\Omega}$ in terms of the multipole operators~\cite{Gir.Bra.Bra:08,Ser.Bra:20,Den.Mar:22}
\begin{equation}
\ket{\Omega}\bra{\Omega} =
\sum_{L=0}^{2j}\sum_{M=-L}^{L} \frac{(2j)! \sqrt{4\pi} \,
 T_{LM}^{\dagger} Y_{LM}(\Omega) }{\sqrt{(2j+L+1)! (2j-L)!}}
\end{equation}
combined with Eqs.~\eqref{rhoTLMexpansion} and \eqref{rho.P.one}. The function $P_0(\rho , \Omega)$ is also obtained by applying the Stratonovich-Weyl map $w^{(s)}(\Omega)$ for $s=1$~\cite{klimov2017gen}, which reads
\begin{equation}
\label{kernel.s}
\begin{aligned}
w^{(1)}(\Omega) ={}& \sqrt{\frac{4\pi}{2j+1}} \sum_{L=0}^{2j} \sum_{M=-L}^L \left( C_{j j L0}^{j j} \right)^{-1} Y_{LM}^{*} (\Omega) T_{LM},
\end{aligned}
\end{equation}
to the state $\rho$
\begin{equation}
\label{kernel.map}
    P_0(\rho , \Omega) = \Tr \big( \rho \, w^{(1)}(\Omega) \big) .
\end{equation}
In Appendix~\ref{App.1}, we show that the eigendecomposition of the kernel $w^{(1)}(\Omega)$ for a general spin $j$ reads
\begin{equation}
\label{Dec.w}
w^{(1)}(\Omega) = \sum_{m=-j}^j
\Delta_{j+m} \ket{j,m ; \Omega}  \bra{j,m ; \Omega} \,
\end{equation}
with
\begin{equation}
\label{Eig.P}
\Delta_{k} =  (-1)^{2j-k} \binom{2j+1}{k} 
\end{equation}
for $k=0, \,  \dots , \, 2j$ and $\ket{j,m ; \Omega} = D(\Omega) \ket{j,m}$.
In particular, its value at the north pole $ (\theta,\phi)=(0,0)$, denoted as $\Omega =0$, reads
\begin{equation}
\label{Gen.P0}
    P_0(\rho , 0) =  \sum_{m=-j}^j
\Delta_{j+m} \bra{j,m } \rho \ket{j,m}\, ,
\end{equation}
As an illustration, we plot its spectrum for small numbers of qubits in Table~\ref{tab:Eig.s1}. The truncated $P_0$ function is a valid $P$ function by itself because it is real, covariant under rotations~\eqref{covariance.P}, and fulfills Eq.~\eqref{rho.P.one.multiqubit}. 
However, it generally does not satisfy the positivity condition, even for separable states. Hence, the additional terms in Eq.~\eqref{general.Pfunction} play a crucial role, as illustrated by examining a spin-coherent state oriented along a specific direction $\Omega_0$. A positive $P$ realization of $\ket{\Omega_0}\bra{\Omega_0}$ is a Dirac delta-function on the sphere, $P(\rho, \Omega) = \delta \left( \Omega - \Omega_0 \right)$. However, it is clear that such a delta function cannot be obtained by truncating the infinite sum in Eq.~\eqref{general.Pfunction}. Indeed, if we did this, the resulting $P$ function would always be negative somewhere on the sphere, as discussed in~\cite{Boh.Gir.Bra:17}.
\subsection{Symmetric absolute separability}
Finally, we introduce the concept of \emph{Symmetric Absolutely Separable} (SAS) states, and denote their set by $\SAS$. SAS states are characterized by the property that every element within their full unitary orbit $\{U\rho U^{\dagger} : U \in SU(2j+1) \}$ is separable. The SAS condition can be reformulated as the existence of a positive $P$ function for $U\rho U^{\dagger}$, that is  
\begin{equation}
\label{general.SAS.cond}
    \min_{\substack{U \in SU(2j+1) \\ \Omega \in S^2}} P(U\rho U^{\dagger} , \Omega) \geqslant 0 \, .
\end{equation}
This formulation implicitly assumes that the $ x_{LM}$ variables in the $P$ function \eqref{general.Pfunction} can depend on the state $\rho$ and the applied unitary transformation $U$.

A \emph{SAS witness}~\footnote{Here, we use the term \emph{witness} for both linear and nonlinear functionals. This generalization has already been used in the literature, see e.g.~\cite{PhysRevLett.96.170502,PhysRevA.76.012334}.}, \ie~a functional of the state $\rho$ that detects some SAS states, can be written only in terms of the eigenvalues of $\rho$ because these form a basis for $SU(2j+1)$ invariant quantities. For example, in the particular case of a two-qubit system, it was shown in Ref.~\cite{ser.mar:23} that a symmetric state $\rho$ with eigenspectrum $(\la_0,\la_1,\la_2)$ sorted in descending order is SAS if and only if
\begin{equation}
\label{SAS.cond.N1}
\sqrt{\la_1} + \sqrt{\la_2} \geq 1 \, .
\end{equation}
For $N=3$, numerical evidence and conjectures about the structure of the set $\SAS$ are also reported in Ref.~\cite{ser.mar:23}. More generally, for an arbitrary $N$, there is a SAS witness derived in Ref.~\cite{Boh.Gir.Bra:17} based on the distance $r$~\eqref{distance.rho.rho0}. It is denoted here by $\WIT_0$, and reads as follows
\begin{witghost}
\label{Witness.0}
$\WIT_0$: A symmetric $N$-qubit state $\rho$ is SAS if its Hilbert-Schmidt distance to the maximally mixed state in the symmetric sector $r$ satisfies
\begin{equation}
\label{rmax.2017}
r^2 \leqslant 
\frac{1}{2(N+1)\left[(2N+1)\binom{2N}{N}-(\tfrac{N}{2}+1)\right]} \, .
\end{equation}  
\end{witghost}
We use the symbol $\Sep_0$ to denote the subset of SAS states witnessed by $\WIT_0$. We use the same notation for subsequent witnesses. 
\begin{table}[t!]
    \centering
    \begin{tabular}{c|c}
         $\;$Number of qubits $N=2j\;$ & $\Delta_{k}$ 
         \\
         \hline 
         2 & $1, \, -3, \, 3$ 
         \\
         3 & $-1 , \, 4 , \, -6, \, 4$
         \\
         4 & $1, \, -5, \, 10, \, -10, \, 5$
         \\
         5 & $-1 , \, 6, \, -15, \, 20, \, -15, \, 6$
         \\
         6 &  $1 , \, -7, \, 21, \, -35, \, 35, \, -21, \, 7$
    \end{tabular}
    \caption{List of eigenvalues $\Delta_{k}$ of $w^{(1)}(\Omega)$ as given by Eq.~\eqref{Eig.P} with $k=0, \, \dots , \, N$ for several number of qubits.} 
    \label{tab:Eig.s1}  
\end{table}
\subsection{Unistochastic and bistochastic matrices}
\label{Subsection.unistochastic}
 We end this section by introducing the concept of $d\times d$ \emph{bistochastic} matrix $B \in \BIS _d$~\cite{marshall2011}. Bistochastic matrices possess positive entries, and the sum of entries in each column or row equals 1. The set of bistochastic matrices, denoted as $\BIS_d$, forms a polytope in $\mathds{R}^{(d-1)^2}$. One approach to parameterise this set involves introducing free variables in a $(d-1) \times (d-1)$ minor of the matrix $B$, and then the remaining entries are determined by satisfying the bistochastic conditions. For instance, $\BIS_3$ can be parametrised by $\rema{\mathbf{b}= (b_1, \, b_2 , \, b_3 \, , b_4)} \in [0, \, 1]^4$ as
\begin{equation}
\label{Par.b.bis}
 B(\mathbf{b}) = \left(
\begin{array}{ccc}
 b_1 & b_2  & 1-b_1 -b_2  \\
 b_3 & b_4  & 1-b_3 -b_4  \\
 1-b_1-b_3 & 1-b_2-b_4  & \sum_{i=1}^4b_i -1  
\end{array}
\right) \, ,
\end{equation}
where the positivity condition of the entries of $B$ defines the domain of the $b_k$ variables~\cite{dun.zcy.09,dita.06}
\begin{align}
\label{BIS.cond}
    \BIS_3 = \Big\{ & B(\mathbf{b}) \big|
    \mathbf{b} \in [0,1]^4 \,  ,
    1 \leq \sum_i b_i , \, b_1+b_2 \leq 1 ,
    \nonumber
    \\
  &  b_1 + b_3 \leq 1, \, b_2+b_4 \leq 1 , \, b_3 + b_4 \leq 1  \Big\}.
\end{align}
For the sake of simplicity, we simply write $\mathbf{b} \in \BIS_3$ whenever $\mathbf{b}$ satisfies these conditions. Another useful parameterization of bistochastic matrices follows from the fact that they are the convex hull of permutation matrices $\si_{\pi}$~\cite{marshall2011}
\begin{equation}\label{Bperm}
    B= \sum_{\pi \in S_d} c_{\pi} \si_{\pi} \, ,
\end{equation}
where $c_{\pi}\geq 0$, $\sum_{\pi} c_{\pi}=1$. 

A special subset of bistochastic matrices are the \emph{unistochastic} matrices $B \in \UNIS_d$ whose entries are specified by a unitary matrix $V \in SU(d)$, $B_{ij} = |V_{ij}|^2$. While for $d=2$, $\UNIS_2 = \BIS_2$, a similar equality does not hold for $d\geq 3$~\cite{dun.zcy.09,dita.06,unistoch.smi}. For example, the vectors $\mathbf{b}$ defining a unistochastic matrix $B\in \UNIS_3$ must satisfy, in addition to the condition $\mathbf{b} \in \BIS_3$, an extra condition related to the positive area of a triangle (see \cite{bengtsson2005birkhoff,dun.zcy.09,dita.06} for more details), such that
\begin{equation}
\label{UNIS.cond}
    \UNIS_3 = \BIS_3 \cap \left\{ 
    B(\mathbf{b}) | A(\mathbf{b})\geq 0
    \right\} \, ,
\end{equation}
with
\begin{equation}
\label{Area.condition}
    A(\mathbf{b}) \equiv 
    4b_1 b_2 b_3 b_4 - \left( b_1 +b_2 + b_3 + b_4 - 1 - b_1 b_4 - b_2 b_3 \right)^2
    \, .
\end{equation}
Note that there is no known full characterisation of unistochastic matrices for $d > 3$~\cite{bengtsson2005birkhoff,unistoch.smi}, \rema{with the exception of partial results, such as the characterisation of circulant matrices for $d=4$~\cite{10.1063/5.0046581} or the study of subsets of bistochastic matrices relevant to quantum information for the general dimension $d$}~\cite{Rajchel.Gasio.Zycz:18}. 
\section{Polytopes of SAS states}
\label{Sec.2.W1}
\begin{figure}[t!]
    \centering
    \includegraphics[width=1.0\linewidth]{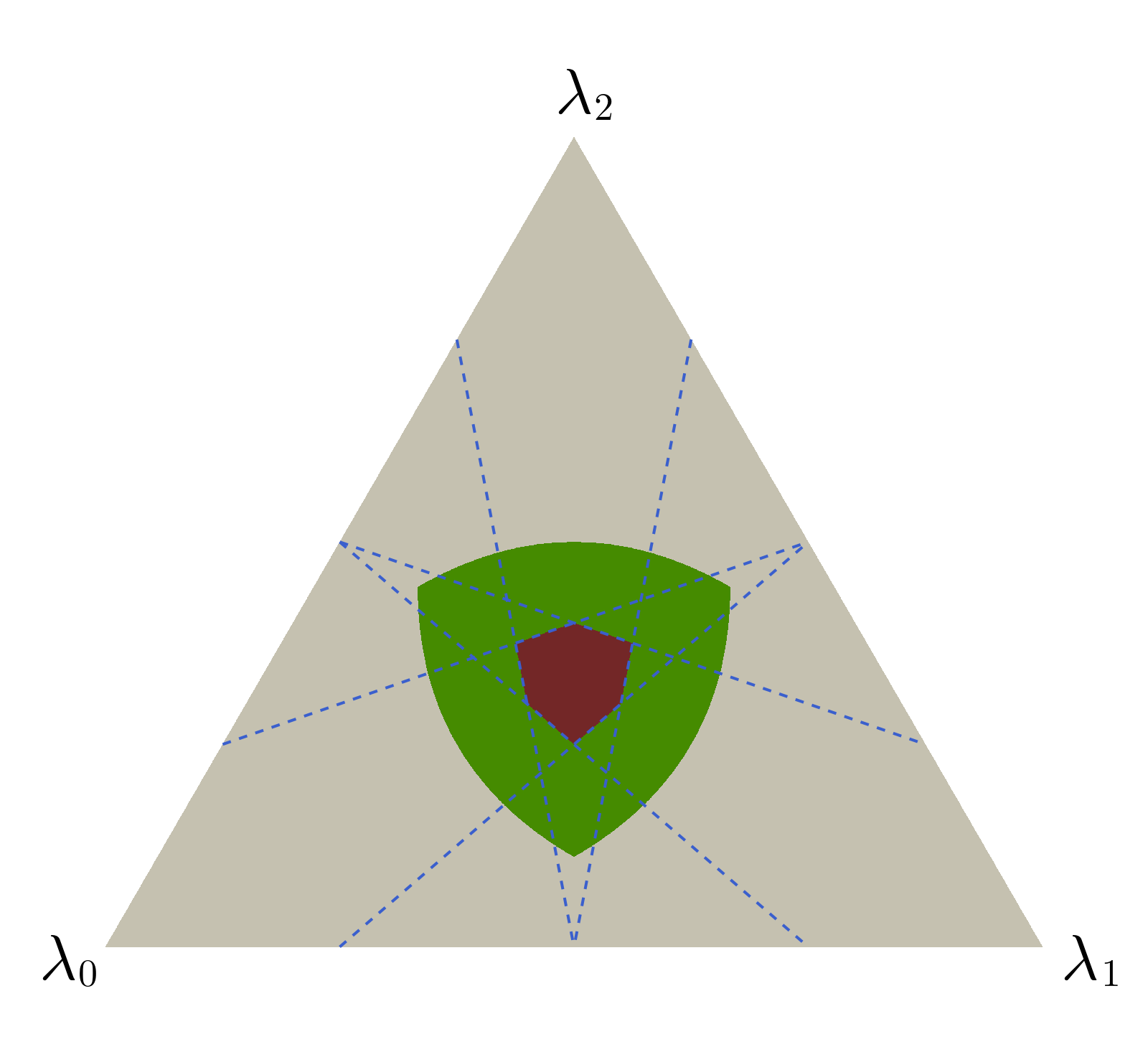}
    \caption{Polytope of SAS states $\Sep_1$ (dark red) for $N=2$ in the simplex of eigenvalues, displayed in barycentric coordinates. The blue dotted lines, whose equations are given by the equality condition in Eq.~\eqref{Per.cond}, define the edges of the polytope. The green area delimited by the condition \eqref{SAS.cond.N1} represents all SAS states (see Ref.~\cite{ser.mar:23}).}
    \label{fig.SAS.Polytope.N2}
\end{figure}
Our first family of SAS witnesses, linear in $\rho$, is obtained by analysing the truncated $P$ function ($P_{\trun}$), \ie {}, Eq.~\eqref{general.Pfunction} with $x_{LM}=0$ for all $L$ and $M$. 
Throughout this section, we use a method similar to that in  Ref.~\cite{denis.martin:23}. First, we write a state in the form $\rho = U \La U^{\da} $, with $\La$ a diagonal matrix in the $\ket{j,m}$ basis with eigenvalues $\la_k$ sorted in non-increasing order. The SAS criterion \eqref{general.SAS.cond} is then translated by the following condition for $P_0$
\begin{equation}
\min_{\substack{U \in SU(2j+1) \\ \Omega \in S^2}} P_{\trun} (U\Lambda U^{\dagger}, \Omega) = \min_{V\in SU(2j+1)} P_0( V \La V^{\dagger} , 0)  \geqslant 0 \, ,
\end{equation} 
where we have used the covariance property of $P_{\trun}$ (see Eq.~\eqref{covariance.P}) to absorb the rotation in the unitary transformation $V\equiv D(\Omega)^{\dagger} U \in SU(2j+1)$. Using Eq.~\eqref{Gen.P0}, the previous expression can be written in terms of the entries $B_{kl} = |V_{kl}|^2$ of a unistochastic matrix $B$ as follows
\begin{equation}
\label{bistoch.mat}
P_0(V\Lambda V^{\dagger} , 0) = \sum_{k , \,l =0}^{2j} \la_k |V_{kl}|^2 \Delta_{l}  = \bm{\la} B \bm{\Delta}^T ,
\end{equation}
where $\bm{\la} = \left( \la_0 , \, \dots , \, \la_{2j} \right)$ is the state eigenspectrum and $\bm{\Delta}= \left( \Delta_0  , \, \dots , \, \Delta_{2j} \right)$ is the kernel eigenspectrum~\eqref{Eig.P}. Using the decomposition of $B$ into permutation matrices
~\eqref{Bperm}, we get
\begin{equation}
P_0 (U\rho U^{\dagger} , \Omega) = \sum_{\alpha=1}^{|S_{2j}|} c_{\alpha} \sum_{k=0}^{2j} \lambda_k \Delta_{\pi_{\alpha}(k)} \, .
\end{equation}
Now, Birkhoff-von Neumann's theorem (see, \emph{e.g.}, Theorem 8.7.2 of Ref.~\cite{hor.joh:12}) establishes that a convex function with respect to the entries $B_{kl}$ has extremal values in the permutation matrices. Moreover, since our function is linear in $B$, the global minimum is then achieved in one of the permutations matrices. In particular, and by following the same line of reasoning as in Ref.~\cite{denis.martin:23}, the SAS condition~\eqref{general.SAS.cond} is met for states whose spectrum satisfies
\begin{equation}
\label{Per.cond}
\sum_{k=0}^{2j}  \lambda_k \Delta_{\pi_{\alpha}(k)} \geqslant 0 
\end{equation}
for each permutation $\pi_{\alpha}$. As explained in Ref.~\cite{denis.martin:23}, all inequalities \eqref{Per.cond} are fulfilled if $\rho$ satisfies the strictest inequality given by 
\begin{equation}
\label{Witness.1.eq}
\bm{\la}^{\downarrow} \bm{\Delta}^{\uparrow \, T} \geqslant   0 \, , 
\end{equation}
where the upper arrow indicates ascending $(\uparrow)$ or descending $(\downarrow)$ eigenvalues sort. In this way, we have derived our first family of SAS witnesses valid for any number of qubits:
\begin{wit}
\label{Witness.1}
 $\WIT_1:$ A symmetric $N$-qubit state $\rho$ is SAS if its eigenspectrum $\bm{\la}$ fulfills $\bm{\la}^{\downarrow } \bm{\Delta}^{\uparrow \, T} \geqslant 0$, where $\Delta_{k} =  (-1)^{N-k} \binom{N+1}{k}$ for $k=0,\ldots,N$.
\end{wit}

\begin{figure}[t!]
    \centering
    \includegraphics[width=1.0\linewidth]{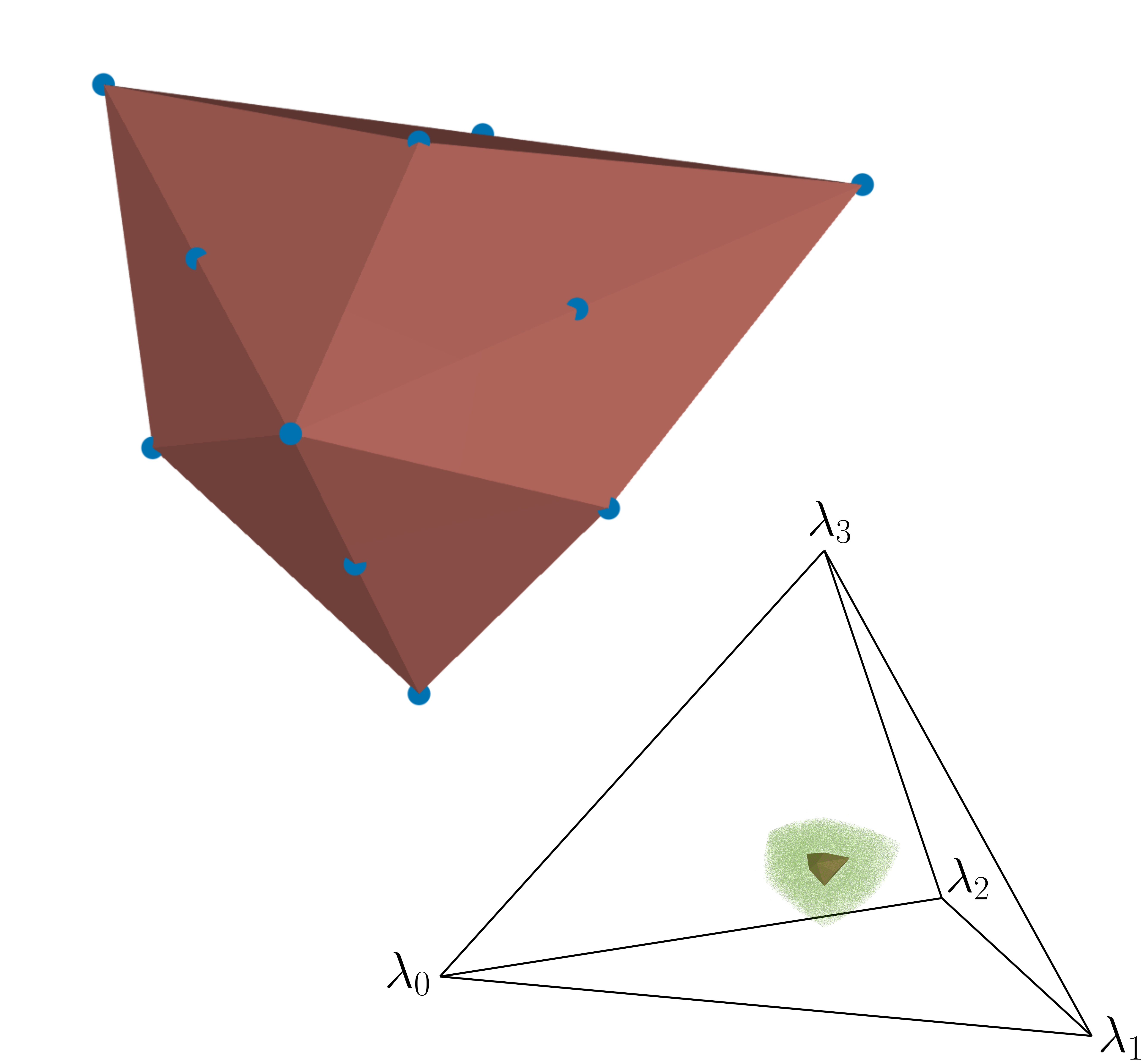}
    \caption{Polytope $\Sep_1$ for $N=3$ next to the full tetrahedron of eigenstates in barycentric coordinates. The vertices of the polytope are indicated by the blue dots. The twice degenerate eigenvalues of $\Delta_k$ (see second row of Table~\ref{tab:Eig.s1}) produce degeneracy in the faces. As a result, some of the blue dots are in the middle of the edges of the polytope. The green points are SAS states obtained by numerical optimization as described in Ref.~\cite{ser.mar:23}.}
    \label{fig.SAS.Polytope.N3}
\end{figure}

The set $\Sep_{1}$ of SAS states detected by the witness $\WIT_1$ typically constitutes a polytope featuring $(N+1)!$ faces, corresponding to the number of inequalities given by~\eqref{Per.cond}, and a total of
\begin{equation}
\sum_{k=1}^{N} \binom{N+1}{k} = 2(2^N-1)
\end{equation}
vertices~\cite{denis.martin:23}, each one given by the intersection of $N$ faces. As an illustration, we plot in Figs.~\ref{fig.SAS.Polytope.N2} and ~\ref{fig.SAS.Polytope.N3} the resulting polytope $\Sep_1$ for $N=2$ and $3$, respectively, in the barycentric coordinate system (see the Appendix of Ref.~\cite{denis.martin:23} for more details on this representation). For $N=2$, $\Sep_{1}$ has 6 faces and 6 vertices according to the previous discussion. However, when dealing with odd values of $N$, the degeneracy in the $\Delta_k$ eigenvalues, as shown in Table~\ref{tab:Eig.s1}, leads to a degeneracy among the faces of $\Sep_1$. Consequently, some vertices are positioned in the middle of an edge. We observe this degeneracy for $N=3$ in Fig.~\ref{fig.SAS.Polytope.N3}, where $\Sep_1$ contains 12 faces instead of 24, and 6 out of the 14 vertices are situated along the middle of certain edges.
\section{Nonlinear SAS witnesses}
\label{Sec.3.W2}
Now that we have characterised all detectable SAS states on the basis of the truncated $P_0$ function, $\Sep_1$, let us consider additional terms of the $P$ function present in the general form~\eqref{general.Pfunction} to derive stronger nonlinear witnesses. In order to simplify the complexity of the minimization problem presented in Eq.~\eqref{general.SAS.cond} and facilitate the derivation of analytical results, we choose to focus exclusively on  additional terms (terms with $L>2j$) that arise from the product of $Q_{L}$ functions, as defined in Eq.~\eqref{Q.function}. Specifically, we consider only those extra terms, denoted hereafter as $P_{L}^{(2j)} = P_{L}^{(2j)}(\rho, \Omega)$, obtained by squaring the $Q_{L}$ functions and then subtracting their lower angular momentum components so that only spherical harmonics with $L>2j$ are involved. We thus add to the truncated $P$ function  terms proportional to
\begin{equation}
\label{ext.terms}
    P_L^{(2j)} \equiv
    Q_{L}^2 - \sum_{\sigma=0}^{2j} \sum_{\nu=-\sigma}^{\sigma} \left( \int Q_{L}^2 \, Y_{\sigma \nu}^* \diff \Omega \right) Y_{\sigma \nu} \, ,
\end{equation}
where only the integrals with $\sigma$ even are non-zero. It is important to note that, by construction, the functions \eqref{ext.terms} are non-zero only for $L > j$ and are covariant under rotations as inherited from the $Q_L$'s. Consequently, we can  work only with the $P$ functions of a unitarily transformed state $\rho'$ evaluated at $\Omega =0$
\begin{equation}
\label{New.P2}
P(U\rho U^{\dagger},\Omega) = P( \rho' , 0) 
    =
    P_{\trun}(\rho' , 0) + \sum_{L > j }^{2j} y_{L} P_{L}^{(2j)}( \rho' , 0) \, ,
\end{equation}
with $\rho' = V\rho V^{\dagger}$, $V= D(\Omega)^{\dagger} U$ and $y_L$ real numbers. Just like the $x_{LM}$ variables in Eq.~\eqref{general.Pfunction}, the $y_L$'s can depend on the state $\rho$ and $U$. However, for the sake of simplicity, we will only consider them as variables independent of the unitary and the state. The general idea for obtaining a SAS witness is to reduce the minimisation problem on the full unitary orbit formulated in Eq.~\eqref{general.SAS.cond} to a problem that requires minimisation on the unistochastic matrices only, as in the developments for obtaining the polytopes of SAS states. 

A direct application of a formula giving the integral of a triple product of spherical harmonics (see e.g.~Eq.~(4), p.~148 of Ref.~\cite{Var.Mos.Khe:88}) together with the equality $|\rho_{L \mu}|=|\rho_{L -\mu}|$ lead us to an algebraic expression for the functions $P_L^{(2j)}(\rho, 0)$ defined in Eq.~\eqref{ext.terms}, 
\begin{equation}\label{PL2j}
 P_L^{(2j)}(\rho,0)  = \frac{2L+1}{4\pi} \sum_{\mu =0}^{L} F(L, \mu ) \left| \rho_{L \mu} \right|^2
\, ,
\end{equation}
with $F(L,\mu)$ state-independent coefficients given by
\begin{equation}
    F(L,\mu) \equiv \left\{   
\begin{aligned}
& 1 - \sum_{\substack{\si=0 \\ \si \text{ even}}}^{2j} \left( C_{L0L0}^{\sigma 0} \right)^2 & \, \, \text{ if } \mu=0 
\\ 
& 2 (-1)^{\mu+1} \sum_{\substack{\si=0 \\ \si \text{ even}}}^{2j} C_{L0L0}^{\sigma 0} C_{L\mu L-\mu}^{\sigma 0}
& \, \, \text{ if } \mu \neq 0
\end{aligned}
\right. \, ,
\label{F.fun}
\end{equation}
where the identity $C_{L0L0}^{\si 0}=0$ for $\si$ odd restricts the sum to $\si$ even. Some values of $F(L,\mu)$ are given in Table~\ref{tab.FLmu}. Note that these coefficients are real and can be positive or negative. The factors $(2L+1)/4\pi$ in Eq.~\eqref{PL2j} can be absorbed into the variables $y_L$ without loss of generality to rewrite Eq.~\eqref{New.P2} as follows
\begin{equation}
\begin{aligned}
\label{New.P2.ver2}
    P(\rho', 0) ={}&  P_0(\rho', 0) + \sum_{L>j}^{2j} \sum_{\mu=0}^L y_L   
F(L,\mu )|\rho'_{L\mu}|^2  \, .
\end{aligned}
\end{equation}
Using Eq.~\eqref{Purity.cond} squared, we can isolate the component $|\rho'_{2j\,1}|^2$ as 
\begin{equation}
\label{clear.rho}
    2|\rho'_{2j\,1}|^2 = {r'}^2 - \sum_{L=1}^{2j} {\rho'_{L0}}^2 - 2 \sum_{L=1}^{2j} \sum_{\mu = 1+\delta_{L,2j}}^L |\rho'_{L \mu}|^2 \, ,
\end{equation}
and then insert this expression into Eq.~\eqref{New.P2.ver2} to get
\begin{equation}
\label{exact.P2}
    \begin{aligned}
        & P(\rho' ,0)= P_0(\rho' ,0) + \Nonli(\rho' ,0) + \tilde{\Nonli}(\rho' ,0)
    \end{aligned}
\end{equation}
with 
\begin{multline}\label{exact.P1def}
\Nonli (\rho' ,0) = \left( \frac{y_{2j} F(2j,1)}{2} \right) {r'}^2 \\ + \sum_{L=1}^{2j} 
        \Big[ y_L F(L,0) \Theta(L- j) - \frac{y_{2j} F(2j,1)}{2} \Big] {\rho'_{L0}}^2
\end{multline}
and
\begin{multline}\label{exact.P2def}
\Tilde{\Nonli} (\rho' ,0) = \sum_{L=1}^{2j} \sum_{\mu = 1+\delta_{L,2j}}^L
        \Big[ y_L F(L,\mu) \Theta(L- j)\\
     - y_{2j} F(2j,1) \Big] \left| \rho'_{L\mu} \right|^2 \, ,
\end{multline}
where $\Theta(x)$ is the Heaviside step-function defined here as
\begin{equation}
    \Theta(x) = \left\{ \begin{array}{cc}
         0 &  x \leq 0 \\
         1 & x > 0
    \end{array} \right. \, .
\end{equation}

{
\begin{table}[t!]
    \centering
    \begin{tabular}{c|l}
         \parbox{3cm}{Number of qubits\\ $N=2j$} & ${\left\{\begin{array}{ll}
         & \mathrm{Witness}~\WIT_1 \\
         & \mathrm{Witness}~\WIT_3
         \end{array}\right .}$\\[10pt]
         \hline \\[-5pt]
         2 &  $\left\{\begin{array}{ll}
         & \bm{\la}\,( -3 , \, 1 , \, 3 )^T \geqslant 0 
         \\[3pt]
           & r^2 \leqslant \frac{1}{78}\approx 0.01282
           \end{array}\right .$
         \\[15pt]
         3 & $\left\{\begin{array}{ll}
         & \bm{\la}\,(-6 , \, -1 , \, 4 ,  \, 4 )^T \geqslant 0 
         \\[3pt]
           & r^2 \leqslant \frac{1}{354}\approx 0.002825
           \end{array}\right .$
         \\[15pt]
         4 & $\left\{\begin{array}{ll}
         & \bm{\la}\,(-10, \, -5, \, 1 , \, 5 , \, 10)^T \geqslant 0 
         \\[3pt]
           & r^2 \leqslant \frac{11}{25390}\approx 0.0004332
           \end{array}\right .$
         \\[15pt]
         5 & $\left\{\begin{array}{ll}
         & \bm{\la}\,(-15, \, -15, \, -1  , \, 6, \, 6 , \, 20)^T \geqslant 0 
         \\[3pt]
           & r^2 \leqslant \frac{1595}{16058598}\approx 0.00009932
           \end{array}\right .$
    \end{tabular}
    \caption{SAS witnesses $\WIT_1$ and $\WIT_3$ for a state with eigenspectrum $\bm{\la} = (\la_0 , \dots , \la_N)$  sorted in descending order $\la_0 \geqslant \la_1 \geqslant \dots \geqslant \la_N$.}
    \label{tab:Witnesses}  
\end{table}
}

In what follows, we will omit the contribution of $\tilde{\Nonli}$ in Eq.~\eqref{exact.P2}, under the assumption that it is positive: $\tilde{\Nonli}(\rho' ,0) \geq 0$. This simplification allows us to reduce the requirement of positivity for $P$ to the simpler condition of positivity for $P_{LB}\equiv P_0+ \Nonli$. The general form of the remaining function $P_{LB}$ reads
\begin{equation}
\label{general.PBL}
P_{LB}(\rho' ) = f(\rho') + \sum_{L=1}^{2j} \left( g_L \rho'_{L0} + h_{L} {\rho'_{L0}}^2 \right)  \, ,
\end{equation}
where the function $f(\rho')$ and the coefficients $g_L$ and $h_L$ are deduced from Eqs.~\eqref{Trun.P0} and \eqref{exact.P2} as being
\begin{equation}
\label{Var.Witness2}
\begin{aligned}
f(\rho') = & \, \frac{1}{2j+1} + \left(  \frac{y_{2j} F(2j,1)}{2} \right) {r'}^2 \, ,
\\
g_L = & \, \sqrt{\frac{2L+1}{2j+1}} \left( C^{jj}_{jj L0} \right)^{-1} \, , 
\\
h_L = & \, y_L F(L,0)\Theta(L- j)   - \frac{y_{2j}F(2j,1)}{2} \, .
\end{aligned}
\end{equation} 
Note that $g_L$ is a constant, $h_L$ depends on the $y_L$ variables, while the function $f(\rho')$ depends only on the $SU(2j+1)$-invariant distance to the maximally mixed state $r'$ given by Eq.~\eqref{distance.rho.rho0}. Hence, $f$, $g_L$ and $h_L$ are constant along the unitary orbit of $\rho'$. 
In the end, the function $P_{LB}$ only depends on the components $\rho_{L0} = \Tr (\rho T_{L0}^\dagger)$ which, for a general element of the unitary orbit $\rho'= V\Lambda V^{\dagger} $, reads
\begin{equation}
\begin{aligned}    
    \rho'_{L0} ={} & \Tr ( V\Lambda V^{\dagger} T_{L0}^\dagger) 
    \\
    = & \sum_{m,m'=-j}^j V_{m'm} \la_m V_{m'm}^* t_{Lm'}
    \\
    ={}& \bm{\la} B\, \mathbf{t}_L^T \,
\end{aligned}
\end{equation}
with $\mathbf{t}_L = (t_{L,j} , , \dots , t_{L,-j})$ the vector containing the eigenvalues of $T_{L0}$~\eqref{decomp.TensOp}, $t_{L,m} = \bra{j,m} T_{L0} \ket{j,m}$, and $B$ the unistochastic matrix defined from $V$, with entries $B_{kl}= |V_{kl}|^2$. The expression of $P_{LB}$ eventually reduces to
\begin{equation}
\label{General.PLB.ver2}
P_{LB}(U \rho U^{\dagger}) = f(\rho') + \sum_{L=1}^{2j} \Big[ g_L \bm{\la} B\, \mathbf{t}_L^T + h_L \left( \bm{\la} B\, \mathbf{t}_L^T \right)^2 \Big].
\end{equation}
As a result, we have moved from a minimization problem over the full unitary orbit of a state to a simpler problem that necessitates a minimization over the unistochastic matrices only, 
\begin{equation}
\label{Seq.inequalities1}
\begin{aligned}
    \min_{\substack{U \in SU(2j+1) \\ \Omega \in S^2}} P(U \rho U^{\dagger} , \Omega) = &     \min_{\substack{V  \in SU(2j+1) \\ V= D(\Rr)^{\dagger} U }} P(V \rho V^{\dagger} , 0) 
\\[0.05cm]
\geqslant & \min_{V \in SU(2j+1)} P_{LB}(V\rho V^{\dagger}) 
\\[0.05cm]
= & \min_{\mathbf{b} \in \UNIS_{2j+1}} P_{LB} (V\rho V^{\dagger})
\end{aligned}
\end{equation}
where in the second line we have used our assumption $\tilde{\Nonli} \geq 0$ and in the last line we refer to a parametrization $\mathbf{b}$ of $\UNIS_{2j+1}$, as discussed in Subsec.~\ref{Subsection.unistochastic}. In particular, the terms $\bm{\la} B \,\mathbf{t}^T_L$ in $P_{LB}$ are linear expressions of the $\mathbf{b}$-variables of the bistochastic matrices. It follows that $P_{LB}$ is quadratic over $\mathbf{b}$ (see also Appendix~\ref{quadratic.minimum}). The inequality \eqref{Seq.inequalities1} is valid as long as the contribution $\tilde{\Nonli}$ given in Eq.~\eqref{exact.P2def} is positive, which defines the admissible range of variation of the parameters $y_L$ in our approach. 

In principle, the minimization on the unistochastic matrices $\UNIS_{2j+1}$ presented in Eq.~\eqref{Seq.inequalities1} seems intractable to perform for $2j=N>3$ due to the lack of a complete characterization of $\UNIS_{2j+1}$. 
However, we can extend the minimization domain to the bistochastic matrices $\mathbf{b} \in \BIS_{2j+1} \supset \UNIS_{2j+1}$ as follows
\begin{equation}
\label{Seq.inequalities2}
\begin{aligned}
    \min_{\substack{U \in SU(2j+1) \\ \Omega \in S^2}} P(U \rho U^{\dagger} , \Omega) 
\geqslant & \min_{\mathbf{b} \in \UNIS_{2j+1}} P_{LB} (V\rho V^{\dagger})
\\
\geqslant & \min_{\mathbf{b} \in \BIS_{2j+1}} P_{LB} (V\rho V^{\dagger}) \, ,
\end{aligned}
\end{equation}
and prove that the second inequality is in fact always tight. This follows from two key facts: i) for any bistochastic matrix $B$, one can always find a unistochastic matrix $B'$ such that $\bm{\la} B = \bm{\la} B'$ for any vector $\bm{\la}$~\footnote{In fact, for any product of a vector and a bistochastic matrix $\bm{\la} B$, one can find an orthostochastic matrix $A$ (\ie~that comes from an orthogonal matrix) such that~$\bm{\la} B = \bm{\la} A $ (See Theorem B.6 of Ref.~\cite{marshall2011}).}, and ii) the bistochastic matrix appears in the objective function~\eqref{General.PLB.ver2} only in the form $\bm{\la} B$. From Eq.~\eqref{Seq.inequalities2}, we can now derive another family of SAS witnesses, denoted as $\WIT_2( \{ y_L \} )$, or just $\WIT_2$ for short, taking the form of a quadratic optimization problem on the $\mathbf{b}$-variables of the bistochastic matrix (see Appendix~\ref{quadratic.minimum}, in particular Eq.~\eqref{quad.problem.PLB}), with the $y_L$ parameters constrained by the positivity condition $\tilde{\Nonli}(\rho',0)\geq 0$~\eqref{exact.P2def}. These witnesses are expressed as follows
\begin{wit}
    \label{Witness.2}
    $\WIT_2 (\{ y_L \} )$: A symmetric $N$-qubit state $\rho$ is SAS if 
\begin{equation}
\label{Witness2}
    \min_{\mathbf{b} \in \BIS_{2j+1}} P_{LB}(U\rho U^{\dagger}) \geqslant 0 \, ,
\end{equation}
where $P_{LB}(U\rho U^{\dagger})$ is given by Eq.~\eqref{General.PLB.ver2} and $y_L$ are real parameters restricted by the inequalities 
\begin{equation}
\label{restriction.ineq}
     y_L F(L,\mu) \Theta(L- j) -  y_{2j} F(2j,1) \geqslant 0 \, ,
\end{equation}
for $L=1, \dots , 2j$ and $\mu = 1 , \dots , L$, where the coefficients $F(L,\mu)$ are defined in Eq.~\eqref{F.fun}.
\end{wit}

\begin{figure}[t!]
    \centering
    \includegraphics[width=0.9\linewidth]{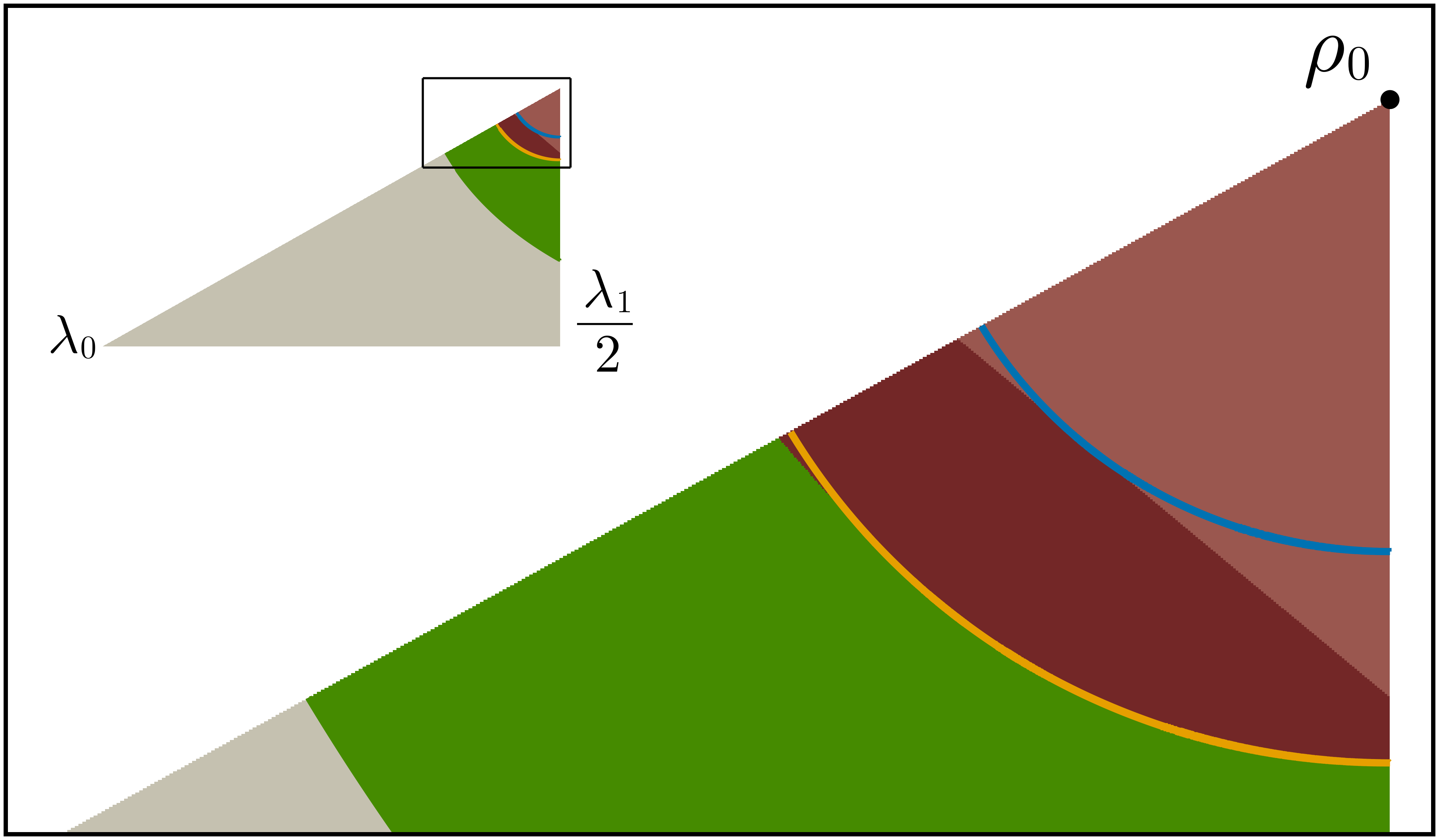}
    \caption{SAS states in the first Weyl chamber ($\lambda_0\geq\lambda_1\geq\lambda_2$) for $N=2$ represented in barycentric coordinates. The curves correspond to the bounds of the sets $\Sep_0$ (blue) and $\Sep_3$ (orange), established by the purity-based witnesses $\WIT_0$ and $\WIT_3$, respectively. The green area encompasses all the remaining SAS states, as completely characterised by the condition \eqref{SAS.cond.N1}. The witnessed states $\Sep_{1}$ are depicted in light red. Lastly, $\Sep_{2}$ with $y_2=455/12$ is formed by the light and dark red regions. We note that $\Sep_{2}$ is only very slightly bigger than $\Sep_{3}$.}
    \label{fig.SAS.N2}
\end{figure}

One can relax the minimization problem to the whole real domain $\mathbf{b} \in \mathds{R}^{4j^2}$ while retaining a useful SAS  witness provided that $h_L > 0$, otherwise the Hessian~\eqref{hessian.B4} of the quadratic function $P_{LB}$~\eqref{General.PLB.ver2} has a negative eigenvalue and hence $P_{LB}$ has a minimum equal to negative infinity.
The latter requirement implies that
\begin{equation}
\label{restrict.hL}
y_L F(L,0)\Theta(L- j) 
>
\frac{y_{2j}F(2j,1)}{2}
\end{equation}
for all $L$. Under these assumptions, the global minimum of $ P_{LB}$ on $\mathbf{b} \in \mathds{R}^{4j^2}$ has an algebraic solution given by (see Appendix~\ref{quadratic.minimum} for details)
\begin{equation}
\label{Global.min}
 \min_{\mathbf{b} \in \UNIS_{2j+1}} P_{LB} 
\geqslant \min_{\mathbf{b} \in \mathds{R}^{4j^2}} P_{LB} = f - \frac{1}{4} \sum_{L=1}^{2j} \frac{g_L^2}{h_L}  \, .
\end{equation}
Any state for which the \rhs~of Eq.~\eqref{Global.min} is positive is then SAS, which after a bit of algebra  can be rewritten as an upper bound on $r$
\begin{equation}
\label{r2.bounds}
    r^2 \leqslant 
    \left( \frac{-2}{y_{2j}F(2j,1)} \right) \left( \frac{1}{2j+1} - \frac{1}{4} \sum_{L=1}^{2j} \frac{g_L^2}{h_L} \right).
\end{equation}
Note that we have used the inequality $y_{2j}\geqslant 0$, which is a direct consequence of $F(2j,1) < 0$ shown in Appendix~\ref{Appendix.Ineq.F} (see Eq.~\eqref{Ineq.1}) and the inequality \eqref{restriction.ineq} for $L=1$. In order to obtain the best SAS witness, we need to maximize the \rhs~of the last equation over the admissible parameters $y_L$. If we first maximize Eq.~\eqref{r2.bounds} with respect to $y_{L}$ for $L\neq 2j$, we find that we need to maximize the $h_L$ variables~\eqref{Var.Witness2} and consequently maximize the $y_L$ variables, which are upper bounded by the conditions~\eqref{restriction.ineq}. Our numerical observations (see Eq.~\eqref{Ineq.5}) show that the strictest upper bound is $y_L \leqslant y_{2j} F(2j,1)/ F(L,1)$. We can then evaluate $y_L$ at these apparently tight upper bounds. Finally, we maximize Eq.~\eqref{r2.bounds} with respect to $y_{2j}$, to obtain the extremal point
\begin{equation}
\begin{aligned}
\label{xM,maximum}
y_L  = & \frac{F(2j,1)}{F(L,1)} y_{2j} 
    \, , \quad \textrm{ for $j < L < 2j$} \, ,
\\
y_{2j} = & \frac{(2j+1)}{F(2j,1)} \sum_{L=1}^{2j} \frac{g_L^2}{ 2 \Theta(L-j) \frac{F(L,0)}{F(L,1)} -1 }
 \, .
\end{aligned}
\end{equation}
In particular, $y_L \geqslant 0$ implies that $h_L > 0$ is fulfilled, as required above. Thus, we obtain a simpler SAS witness $\WIT_3$, weaker than $\WIT_2$, but with an analytical expression:
\begin{wit}
\label{Witness.3}
$\WIT_3$: A symmetric $N$-qubit state $\rho$ is SAS if its distance $r$ to the maximally mixed state in the symmetric sector fulfills
\begin{equation}
\label{witness.distance}
r^2 \leqslant \frac{1}{(2j+1)^2} \left( \sum_{L=1}^{2j} \frac{g_L^2}{1- 2 \Theta(L-j) \frac{F(L,0)}{F(L,1)}  } \right)^{-1}
\end{equation}
with $F(L,\mu)$ and $g_L$ state-independent constants defined in Eqs.~\eqref{F.fun} and \eqref{Var.Witness2}.
\end{wit}
One final remark needs to be made about $\WIT_3$. We can eliminate another $\rho_{L\mu}$ in Eq.~\eqref{clear.rho} instead of $\rho_{2j\, 1}$. However, the witnessed SAS set does not change substantially.
\section{Two- and three-qubit cases}
\label{Sec.5}
In this section, we exemplify our witnesses in the specific cases of $N=2$ and $N=3$, in order to clarify all the technical aspects of our method.
\subsection{\texorpdfstring{$N=2$ qubits}{Lg}}
In this case, the witness $\WIT_1$ is given by 
\begin{equation}
    \bm{\lambda}^{(\uparrow)} (-3 ,1 ,3)^{T} \geqslant 0 \, .
\end{equation}
We plot in Figs.~\ref{fig.SAS.Polytope.N2} and \ref{fig.SAS.N2} the witnessed set $\Sep_{1}$ in the whole eigenspectra simplex of the state and in one Weyl chamber, respectively. In contrast, the other two witnesses, $\WIT_2 (\{ y_L \})$ and $\WIT_3$, rely on the $P$ function~\eqref{New.P2} with only one extra term
\begin{equation}\label{PN2}
    P(\rho,\Omega) = P_0(\rho,\Omega)  + y_2 P_2^{(2)} (\rho,\Omega) \, ,
\end{equation}
with 
\begin{equation}
\label{P2_2}
  \frac{4\pi}{5} P_{2}^{(2)}(\rho, 0)
    = \frac{6}{35} \left( 3  \rho_{20}^2 - 4 | \rho_{21} |^2 + |\rho_{22}|^2 \right) \, .
\end{equation}
By scaling $y_2$ with the factor $5/4\pi$, and performing the change of variable~\eqref{clear.rho} in Eqs.~\eqref{PN2}-\eqref{P2_2}, reading explicitly
\begin{equation}
    2|\rho_{21}|^2 = r^2 - \sum_{L=1}^2 \rho_{L0}^2 - 2 \sum_{
        L=1}^2 \sum_{M=1+\delta_{L,2}}^L |\rho_{LM}|^2  \, ,
\end{equation}
we get
\begin{equation}
\label{W2.N2}
\begin{aligned}
P(\rho,0) ={}& \, P_0(\rho,0) + \Nonli(\rho,0) + \tilde{\Nonli}(\rho,0) 
\, ,
\end{aligned}
\end{equation}
with
\begin{equation}
\begin{aligned}
 \Nonli (\rho,0) ={} &   \frac{6}{35}y_2 \Big( -2r^2  + 2\rho_{10}^2 + 5\rho_{20}^2 \Big) \, ,
\\
 \tilde{\Nonli} (\rho,0) ={} & \frac{6}{35}y_2 \Big(   4|\rho_{11}|^2+ 5 |\rho_{22}|^2
\Big) \, .
\end{aligned}
\end{equation}
The inequality $P \geqslant P_0 + \Nonli \equiv P_{LB}$ defines the admissible region~\eqref{restriction.ineq} of the $y_2$ variable, here $y_2 \geqslant 0$. The function $P_{LB}$ has now the form \eqref{general.PBL} with factors~\eqref{Var.Witness2} given by
\begin{equation}
\label{Coef.W2.N2}
\begin{aligned}
& f(\rho) = \frac{1}{3} - \frac{12}{35}y_2 r^2 \, ,  \\
& (g_1, g_2 )=\left( \sqrt{2} , 5\sqrt{\frac{2}{3}} \right) \, ,\\
& (h_1 , h_2 ) = \frac{6}{35} (2y_2,5y_2 ) \, ,
\end{aligned}
\end{equation}
which are the elements needed to calculate the witness $\WIT_2(\{ y_L \})$ defined by Eq.~\eqref{Witness2}. Lastly, we use Eq.~\eqref{witness.distance} to calculate our third witness at $y_2 = 455/12$~\eqref{xM,maximum} 
\begin{equation}
\WIT_3~\mathrm{for~2~qubits}:\; \rho \in \SAS \, \text{ if } \,  r^2 \leqslant \frac{1}{78} \, .
\end{equation}
\begin{figure}[t!]
    \centering  \includegraphics[width=0.9\linewidth]{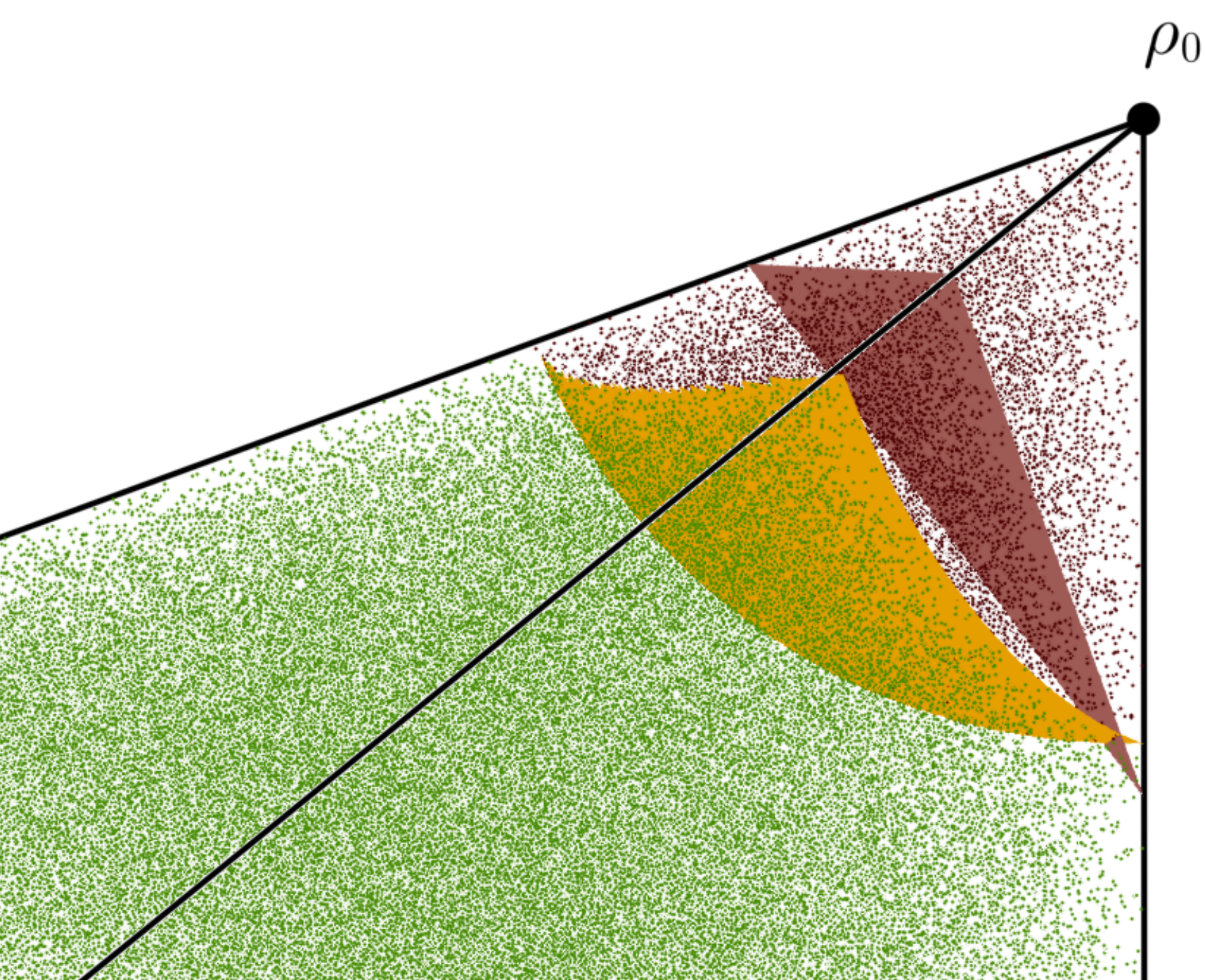}
    \caption{SAS states in the first Weyl chamber ($\lambda_0\geq\lambda_1\geq\lambda_2\geq\lambda_3$) for $N=3$. The dark red and orange surfaces are the boundaries of $\Sep_{1}$ and $\Sep_{3}$, respectively. The dark red points are the states contained in $\Sep_{2}(\{y_L\})$ with $y_L$ equal to Eq.~\eqref{xM,maximum}. The green points are SAS states obtained numerically as described in Ref.~\cite{ser.mar:23}.}
    \label{fig.SAS.N3}
\end{figure}
\subsection{\texorpdfstring{$N=3$ qubits}{Lg}}
The linear witness $\WIT_1 $ is given by 
\begin{equation}
\bm{\lambda}^{(\uparrow)} (-6 ,-1 ,4,4)^{T} \geqslant 0 \, .
\end{equation}
Its corresponding witnessed SAS set $\Sep_1$ is plotted in Figs.~\ref{fig.SAS.Polytope.N3} and \ref{fig.SAS.N3} in dark red. Now, for the nonlinear witnesses, we start by defining the respective $P$ function~\eqref{New.P2} which has  two additional terms
\begin{equation}
\label{Pfunction.N3}
    P = P_0 + y_2 P_2^{(3)}+ y_3 P_3^{(3)} \, ,
\end{equation}
where $P_2^{(3)} = P_2^{(2)}$ [given by Eq.~\eqref{P2_2}] and 
\begin{equation}
   \frac{4\pi}{7}
   P_3^{(3)}(\rho, 0)
    = \frac{2}{21} \left( 7\rho_{30}^2 - 6 |\rho_{31}|^2 - 3 |\rho_{32}|^2 + 2 |\rho_{33}|^2  \right)
    \, .
\end{equation}
By using Eq.~\eqref{clear.rho} to substitute $|\rho_{31}|^2$ in the $P$ function, we get $P = P_0 + \Pi + \tilde{\Pi}$ with
\begin{equation}
\begin{aligned}    
    \Pi  =& \frac{2}{7} \Bigg[ x_3 \left(  \rho_{10}^2 -r^2 \right) + \left( \frac{9y_2+5y_3}{5} \right) \rho_{20}^2
    + \frac{10}{3} y_3 \rho_{30}^2 \Bigg] \, ,
    \\
    \tilde{\Pi} = & \frac{2}{7} \Bigg[ 2y_3 |\rho_{11}|^2 + \frac{2}{5} \left( 5y_3-6y_2 \right) |\rho_{21}|^2  
    \\
    & + \left( \frac{3}{5} y_2 + 2y_3 \right) |\rho_{22}|^2 +y_3 |\rho_{32}|^2 + \frac{8}{3} y_3 |\rho_{33}|^2 \Bigg] \, .
    \end{aligned}
\end{equation}
Again, $P \geqslant P_0 + \Pi \equiv P_{LB}$ when all the coefficients of the terms in $\Tilde{\Pi}$ are positive, \ie,
\begin{equation}
\label{N3.Witbis.ine}
y_3 \geqslant 0 \, , \quad 5y_3 - 6y_2 \geqslant 0
\, , \quad \frac{3}{5}y_2 + 2y_3 \geqslant 0 
 \, .
\end{equation} 
Let us remark that $y_2$ may take negative values. 
Now, we can calculate the factors~\eqref{Var.Witness2} which are necessary to calculate $P_{LB}$ and the witness $\WIT_2 (\{ y_L \} )$. They are given by
\begin{equation*}
 f(\rho) = \frac{1}{4} - \frac{2}{7} y_3 r^2 , \, \quad (g_1, g_2 ,g_3) = \left( \frac{\sqrt{5}}{2} , \frac{5}{2} , \frac{7\sqrt{5}}{2} \right) \, , 
\end{equation*}
\begin{equation}
 (h_1, h_2, h_3) = \frac{2}{7} \left(  y_3 , \frac{9}{5}y_2+y_3 , \frac{10}{3}y_3
 \right)  \, ,
\end{equation}
Finally, for $\WIT_3$, we have the extra condition $h_L > 0$~\eqref{restrict.hL} for the $y_L$ values
\begin{equation}
    9y_2 + 5y_3 > 0 \, ,
\end{equation}
which is always satisfied for the extremal $y_L$ values given in Eq.~\eqref{xM,maximum}, which read
\begin{equation}
y_2 = \frac{2065}{16} \, , \quad y_3 = \frac{1239}{8} \, .
\end{equation}
The witness $\WIT_3$ eventually reads 
\begin{equation}
\WIT_3~\mathrm{for~3~qubits}:\; \rho \in \SAS \; \text{ if } \; 
 r^2 \leqslant \frac{1}{354} \, .  
\end{equation}
\section{Discussion and comparison between SAS witnesses}
\label{Sec.4.disc}
\begin{figure}[t!]
    \centering
    \includegraphics[width=1.0\linewidth]{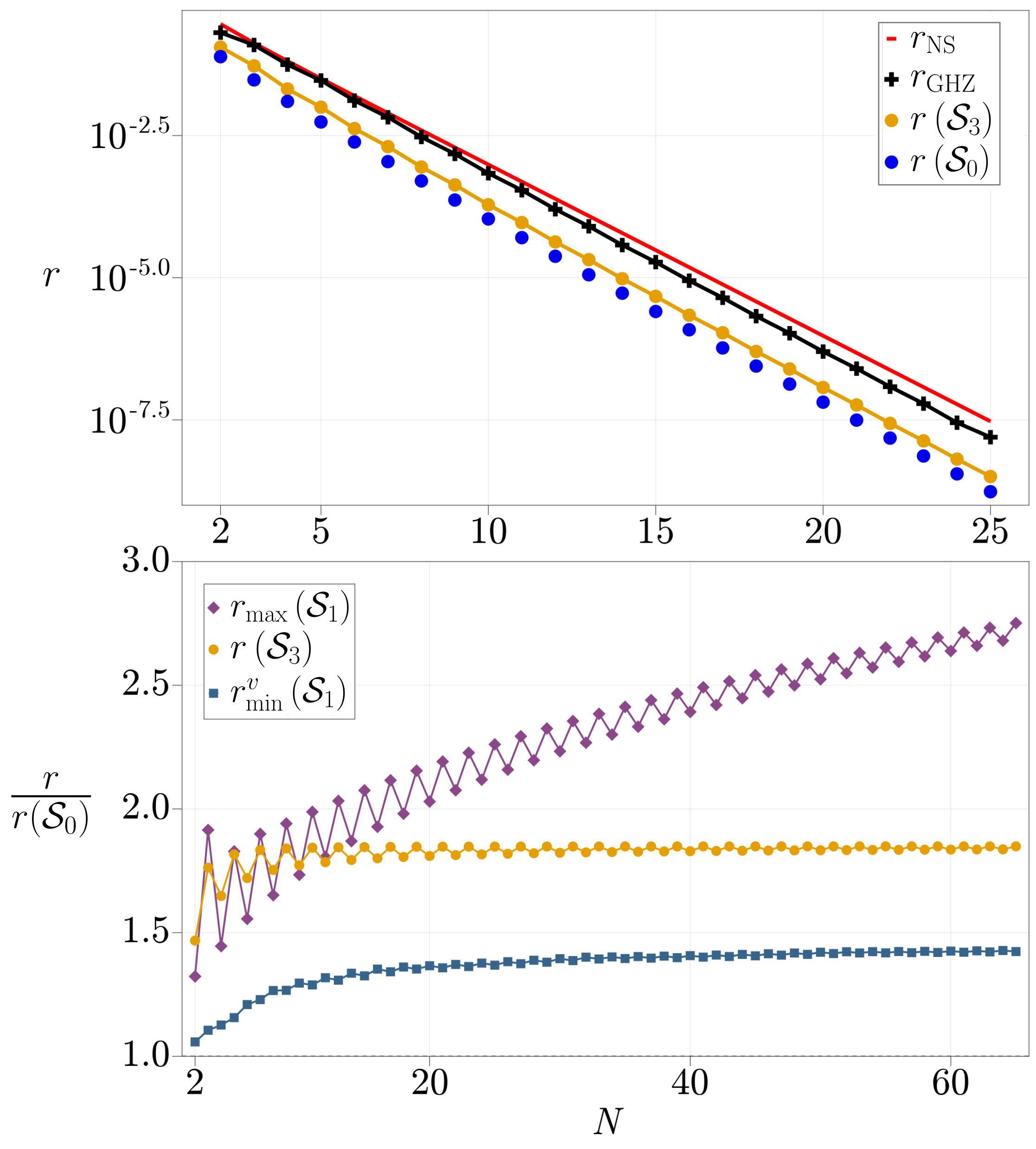}
    \caption{
    Top: Distances $\dist(\Sep_0)$ (blue) and $\dist(\Sep_3)$ (orange) as defined in Eq.~\eqref{Dist.SAS}. The black crosses show $r_{\mathrm{GHZ}}$, defined by the NPT entangled state of the form~\eqref{rho.GHZ} closest to $\rho_0$, which provides an upper bound to the radius of the largest inner ball contained in $\SAS$. The red line shows the radius $r_{\mathrm{NS}}$ of the largest ball containing only AS states in the full Hilbert space (see Eq.~\eqref{r.ns}). Bottom: Distances $\dist_{\max} (\Sep_1)$ (purple), $\dist^v_{\min} (\Sep_1)$ (blue) and $\dist (\Sep_3)$ (orange), rescaled by the distance of the witness $\dist (\Sep_0)$, as a function of the number of qubits. $\dist_{\min}^v (\Sep_1)$ corresponds to the minimal distance between $\rho_0$ and the vertices of the polytope $\Sep_{1}$. Note that all ratios are larger than 1, indicating an improvement on previous results.}
    \label{Fig.Comparison_radii}
\end{figure}
Let us now discuss the physical implications of our SAS witnesses ($\WIT_1$, $\WIT_2$ and $\WIT_3$) and explore their differences. In Table~\ref{tab:Witnesses}, we present the  witnesses $\WIT_1$ and $\WIT_3$ for $N$ ranging from 2 to 5, while the witness $\WIT_2(\{ y_L \})$ generally requires a numerical optimization. To highlight the differences between these witnesses, we illustrate their corresponding sets $\Sep_{k}$ in Figs.~\ref{fig.SAS.Polytope.N2}-\ref{fig.SAS.N3} for $N=2$ and $N=3$, respectively, alongside the set $\Sep_0$ defined by the witness $\WIT_0$ previously obtained in Ref.~\cite{Boh.Gir.Bra:17}. In addition, we provide a supplementary video~\cite{Supp.mat:23.Pfunc} showing $\Sep_{2}(\{ y_L\})$ as $y_L$ varies for $N=2$. 
Below, we give some general observations and remarks about our witnesses.

The sets $\Sep_{k}$ exhibit diverse geometric forms. $\Sep_{1}$ is represented as a polytope, in contrast to $\Sep_0$ and $\Sep_{3}$, which appear as balls centered around $\rho_0$. Furthermore, $\Sep_{2}$ generally exhibits a more complex and non-trivial shape. By construction, $\Sep_{1}=\Sep_{2}(\{y_L =0\})$ and $\Sep_3 \subset \Sep_{2}(\{y_L \})$ for $\{ y_L \}$ given by Eq.~\eqref{xM,maximum}. A detailed analysis reveals that the ball $\Sep_0$ is included in and tangent to $\Sep_1$, meaning that the largest inner SAS ball contained within $\Sep_1$ actually coincides with $\Sep_0$. We prove this result for all values of $N$ in Appendix~\ref{App.B}.

For $N=2$ and $N=3$, we observe that $\Sep_0 \subsetneq \Sep_k$ for $k=1, 2, 3$. This result is expected since $\Sep_0$ is tangent to $\Sep_1$, $\Sep_3$ is maximized such that it witnesses a bigger SAS ball than $\Sep_0$, and $\Sep_3 \subset \Sep_2(\{y_L\})$ for the $y_L$ values~\eqref{xM,maximum}. We can appreciate this behaviour in Fig.~\ref{Fig.Comparison_radii} where we plot the distances
\begin{equation}
        \begin{aligned}
        \label{Dist.SAS}
            \dist_{\max}(\Sep_k) & \equiv \max_{\rho \in \partial \Sep_k}  \big\lVert \rho - \rho_0 \big\rVert_{\text{HS}} =   
            \max_{\rho \in \Sep_k}  \big\lVert \rho - \rho_0 \big\rVert_{\text{HS}}
            \, ,
            \\
            \dist_{\min}(\Sep_k) & \equiv \min_{\rho \in \partial \Sep_k}  \big\lVert \rho - \rho_0 \big\rVert_{\text{HS}} \, ,
        \end{aligned}
    \end{equation}
where $\partial \Sep_k$ is the boundary of $\Sep_k$. In particular, 
    \begin{equation}
        \dist_{\max}(\Sep_0) = \dist_{\min}(\Sep_0) \equiv \dist(\Sep_0) \, ,
    \end{equation}
and the same holds for $\Sep_3$. Specifically, we can observe in Fig.~\ref{Fig.Comparison_radii} that $\dist(\Sep_0) \leqslant  \dist_{\max}(\Sep_1) , \, \dist(\Sep_3) $. By the previous remark, we also have that $\dist_{\min}(\Sep_1) = \dist(\Sep_0)$. The discrepancy between the SAS states identified by $\Sep_2(\{ y_L \} )$ and $\Sep_3$ is notably small for $N=2$ when $y_L$ is determined by Eq.~\eqref{xM,maximum} (see Fig.~\ref{fig.SAS.N2}). This difference is even less pronounced for $N=3$, making it difficult to discern visually.

For $N=2$, $\Sep_1$ is a proper subset of $\Sep_3$, while for $N=3$, the sets identified by the two witnesses are complementary. We anticipate a similar behaviour for larger values of $N$, as confirmed by Fig.~\ref{Fig.Comparison_radii} showing that $\dist (\Sep_0)= \dist (\Sep_1) < \dist (\Sep_3) < \dist_{\max} (\Sep_1)$ from $N=11$ to $65$. A noteworthy consequence of the previous observation is that the furthest away vertex of $\Sep_1$ from $\rho_0$ is not contained in $\Sep_3$. Conversely, the closest vertex of $\Sep_1$ to $\rho_0$ is included in $\Sep_3$, a characteristic that has been observed, as depicted in Fig.~\ref{Fig.Comparison_radii}, where we present the minimal distance between $\rho_0$ and the vertices of the polytope $\WIT_1$, denoted as $\dist^v_{\min} (\Sep_1)$. The values of $\dist(\Sep_3)/\dist (\Sep_0)$ and $\dist^v_{\min} (\Sep_1) / \dist (\Sep_0)$ reach asymptotic values, observed at around 1.849 and 1.423, respectively.

Finally, we stress once again that our witnesses do not detect the whole $\SAS$ set. To get a better idea of its size, we can establish an upper bound on the radius of the largest ball contained within $\SAS$, denoted by $r_{\min}(\SAS)$. Such a bound can be obtained by computing for example the distance $r_{\mathrm{GHZ}}$ above which the mixture of a pure GHZ state and $\rho_0$,
\begin{equation}
    \label{rho.GHZ}
\rho_p=p|\mathrm{GHZ}\rangle\langle \mathrm{GHZ}|+(1-p)\rho_0,
\end{equation}
becomes NPT, i.e.\ has a negative partial transpose and thus entanglement that can be detected by the PPT criterion. The resulting upper bound, which is not tight but improves the numerical bound found in Ref.~\cite{Boh.Gir.Bra:17}, is plotted in Fig.~\ref{Fig.Comparison_radii} along with the radii of the other witnessed SAS sets $\Sep_0$ and $\Sep_3$. The radius of the largest inner ball in $\SAS$ lies between the yellow and the black lines of Fig.~\ref{Fig.Comparison_radii}. For comparison, we also plot the radius of the largest ball containing only AS states in non-symmetric multiqubit systems~\cite{Gur:02}
\begin{equation}
\label{r.ns}
r_{\mathrm{NS}}= \frac{1}{\sqrt{2^N(
2^N -1)}} \, .    
\end{equation}
\section{Conclusions and perspectives}
\label{Sec.5.Con}
In this work, we have derived three families of SAS witnesses for mixed symmetric multiqubit states, one linear ($\WIT_1$) and two nonlinear ($\WIT_2$ and $\WIT_3$) in the \rema{eigenvalues of the} state. Each of these witnesses detects a greater number of SAS states than the witness $\WIT_0$ previously established in Ref.~\cite{Boh.Gir.Bra:17}. We have thoroughly explored their distinctions and delved into their geometric properties. Particularly, we showed that the set $\Sep_1$ of SAS states detected by $\WIT_1$ is a polytope that encompasses and is tangent to the SAS ball defined by $\WIT_0$. To get the formal proof of this result, we derived an analytical expression for the eigenvalues of the Stratonovich-Weyl kernel $w^{(1)}(\Omega)$~\eqref{Eig.P}. This expression can be useful beyond this work, in contexts where quantum states are examined in their phase space (see Ref.~\cite{klimov2017gen} for a review). Furthermore, all the SAS witnesses introduced in this work can be applied to detect absolute classicality~\cite{Boh.Gir.Bra:17} in the framework of spin-$j$ systems via the correspondence explained in Sec.~\ref{concepts.Sec}. Among our witnesses, $\WIT_1$ and $\WIT_3$ are simple functionals on the spectrum and the purity of the state. In contrast, $\WIT_2(\{y_L\})$, which is capable of detecting a larger subset of SAS states, necessitates solving a quadratic optimization problem over the bistochastic matrices. Nevertheless, our observations on the cases $N=2$ and $N=3$ indicate that the disparity between the SAS subsets detected by $\WIT_3$ compared to $\WIT_2$ is minimal, as illustrated in Fig.~\ref{fig.SAS.N2}. It is also important to note that none of our witnesses provides complete detection of all SAS states, i.e.\ the full~$\SAS$ set. This limitation highlights the need for further exploration towards complete detection of SAS states. To efficiently uncover more SAS states, instead of scrutinizing $\WIT_2$, a viable approach is to include additional terms in the $P$ function, as outlined in Eq.~\eqref{ext.terms}. The minimum number of terms to be added for a complete characterisation remains as an open question, even in the case of two qubits. 

Another strategy for witnessing larger sets of SAS states is to use the convexity property of $\SAS$~\footnote{The proof of the convexity of the set of symmetric absolutely separable states is a special case of the convexity of the absolutely separable case~\cite{PhysRevA.89.052304}.}. In particular, we can immediately establish a stronger SAS witness defined by the convex hull of all $\Sep_k$ found in this work. Moreover, as $N$ increases, the discrepancy between $\Sep_1$ and $\Sep_3$ (see Fig.~\ref{Fig.Comparison_radii}) and, subsequently, the convex union of the sets, becomes larger, as does the number of SAS states detected. \rema{The convexity of $\SAS$ in the eigenspectra simplex could be also exploited to define a notion of optimal SAS witness that depends linearly on the spectrum of the state. This would be the analogous concept of optimal linear witnesses for entanglement~\cite{GUHNE20091}.} In terms of experimental implementation, measuring the state spectrum is required for $\WIT_1$ and $\WIT_2$, while $\WIT_3$ relies on evaluating the state's purity. Importantly, both the spectrum and purity of a quantum state can be estimated without the need for full quantum tomography~\cite{PhysRevLett.88.217901}, which facilitates the implementation of our witnesses.

\rema{Lastly, it should be mentioned that there is a debate in the literature about whether entanglement in systems of indistinguishable particles can arise artificially from their exchange symmetry. Consequently, alternative definitions of entanglement have been introduced that extend the set of separable states to include those states whose entanglement comes solely from
exchange symmetry, see e.g.~\cite{ghirardi2002entanglement}. Nevertheless, it is interesting to note in this context that if a state is not entangled in the usual sense of the term, neither will it be entangled in the alternative definitions of entanglement. Consequently, the SAS states detected for the witnesses $\WIT_1, \WIT_2$ and $\WIT_3$ are also SAS for alternative definitions of entanglement.
}
\section*{Acknowledgements}
ESE acknowledges support from the postdoctoral fellowship of the IPD-STEMA program of the University of Liège (Belgium). Computational resources were provided by the Consortium des Equipements de Calcul Intensif (CECI), funded by the Fonds de la Recherche Scientifique de Belgique (F.R.S.-FNRS) under Grant No. 2.5020.11. Most of the computations were done with the Julia programming language. \rema{The authors thank Grzegorz Rajchel for his fruitful correspondence.}
\begin{appendix}
\section{Eigendecomposition of the Stratonovich-Weyl map}
\label{App.1}
%
Here we derive the eigendecomposition of  Eqs.~\eqref{Dec.w}-\eqref{Eig.P} of $w^{(1)}(\Omega)$. We start with the expression of the Stratonovich-Weyl map $w^{(1)}(\Omega)$~\eqref{kernel.s} evaluated at $\Omega=0$
\begin{equation}
\begin{aligned}
& w^{(1)}(0) = \sqrt{\frac{4\pi}{2j+1}} \sum_{L=0}^{2j} \sum_{M=-L}^L \frac{ Y_{LM}^{*} (0) T_{LM} }{C_{j j L0}^{j j}}
\\
& =
\sum_{m=-j}^j \Bigg[ \frac{(-1)^{j-m}}{\sqrt{2j+1}} \sum_{L=0}^{2j}  \frac{\sqrt{2L+1} C_{jm j-m}^{L0}}{C_{j j L0}^{j j}} 
 \Bigg] \ket{j,m}\bra{j,m}
\\
& = \sum_{m=-j}^j \Delta_{j+m} \ket{j,m}\bra{j,m}
\, ,
\end{aligned}
\end{equation}
where we used Eq.~\eqref{decomp.TensOp} for the multipole operators. The operator $w^{(1)}(\Omega)$ is covariant under rotations~\cite{klimov2017gen}, \ie, 
\begin{equation}
    D(\Omega) w^{(1)}(0) D(\Rr)^{\dagger}(\Omega) = w^{(1)}(\Omega) \, .
\end{equation}
Therefore, the eigenvectors of $w^{(1)}(\Omega)$ are the standard spin eigenstates over the quantization axis in the direction $\Omega$, $\ket{j,m ; \Omega} = D(\Omega) \ket{j,m} $. On the other hand, the eigenvalues $\Delta_{j+m}$ can be rewritten as
\begin{equation}\label{Deltajpm}
\begin{aligned}
\Delta_{j+ m}  
=  &
\sum_{L=0}^{2j} \frac{(-1)^{j-m}}{(2j+1)!}  C_{jm,j-m}^{L0} 
\\
& \times \sqrt{(2L+1)(2j-L)!(2j+L+1)!}
\, ,
\end{aligned}
\end{equation}
by using the identity~\cite{Var.Mos.Khe:88}
\begin{equation}
\label{eq:CB_relations}
C_{jj L0}^{jj} = (2j)! \sqrt{\frac{ 2j+1}{(2j-L)!(2j+L+1)!} } \, .
\end{equation}
In order to evaluate the sum in \eqref{Deltajpm}, our strategy is to write a polynomial whose coefficients are proportional to the $\Delta_{j+m}$ eigenvalues deduced above. We start with a generating function of the Cleabsch-Gordan coefficients in terms of the Legendre polynomials, given through the hypergeometric function $_2F_1(a_1, \, a_2; \, b_1; \, t)$ (see Eq.~(5), p.~263 of Ref.~\cite{Var.Mos.Khe:88} or Ref.~\cite{akim1961}, and Eq.~(4), pp.~976 and 1005 of Ref.~\cite{gradshteyn2014table}))
\begin{multline}\label{CL0jmjmm}
\sqrt{\frac{2L+1}{(2j-L)!(2j+L+1)!}} (t-1)^{2j}
P_L \left(\frac{t+1}{t-1} \right)
\\
= \sqrt{\frac{2L+1}{(2j-L)!(2j+L+1)!}} (t-1)^{2j-L} {}_2 F_1(-L, \, -L ; \, 1 ; \, t) 
\\
=
\frac{1}{(2j)!} 
\sum_{m=-j}^j \binom{2j}{j+m} C_{jm, \, j-m}^{L 0} t^{j+m} 
\, .
\end{multline}
We now multiply Eq.~\eqref{CL0jmjmm} by the necessary factors and sum over $L$ to bring out the eigenvalues $\Delta_{j + m}$
\begin{equation}
\label{App.Eq0}
\begin{aligned}
& (t-1)^{2j} \sum_{L=0}^{2j} \left(\frac{2L+1}{2j+1} \right)  P_L \left(\frac{t+1}{t-1} \right)
\\
 &{}=
\sum_{m=-j}^j \binom{2j}{j+m} (-1)^{-j +m}  \Delta_{j+m} t^{j +m}
\\
&{}=
\sum_{k=0}^{2j} \binom{2j}{k} (-1)^{-2j+k} \Delta_{k} t^{k} 
\, .
\end{aligned}
\end{equation}
Now, we can rewrite the left hand-side of Eq.~\eqref{App.Eq0} as
\begin{equation}
\begin{aligned}
& \sum_{L=0}^{2j} \left(\frac{2L+1 }{2j+1}\right) P_L \left( \frac{t+1}{t-1} \right) 
\\
= {} & \frac{t-1}{2} \left[ P_{2j+1} \left(\frac{t+1}{t-1} \right) - P_{2j} \left(\frac{t+1}{t-1} \right) \right] \, ,
\end{aligned} 
\end{equation}
where we used the Christoffel's identity (see e.g.~Eq.~(1), p.~986 of Ref.~\cite{gradshteyn2014table}) with $x=\frac{t+1}{t-1}$, $y=1$ and $n=2j$, that is
\begin{multline}
 \sum_{k=0}^n \left( \frac{2k+1}{n+1} \right) P_k(x) P_k(y) 
    =\\ \frac{ P_{n+1}(x) P_n(y) - P_{n}(x) P_{n+1}(y) }{x-y} \, .
\end{multline}
Lastly, we use the power expansion of the Legendre polynomials~\cite{gradshteyn2014table}
\begin{equation}
\label{Hyp.Leg}
(t-1)^L P_L \left(\frac{t+1}{t-1} \right)
= 
\sum_{a=0}^L \binom{L}{a}^2 t^a
\end{equation}
to get
\begin{equation}
\label{App.Eq1}
\begin{aligned}
& (t-1)^{2j} \sum_{L=0}^{2j} \left(\frac{2L+1}{2j+1} \right)  P_L \left(\frac{t+1}{t-1} \right)
\\
={}& 
\frac{1}{2} \left[ 
\sum_{k=0}^{2j+1} \binom{2j+1}{k}^2 t^k - (t-1) \sum_{k=0}^{2j} \binom{2j}{k}^2 t^k
\right]
\\
 ={}& 
 1 + \sum_{k=1}^{2j} \left[ 
\binom{2j+1}{k}^2 + \binom{2j}{k}^2 - \binom{2j}{k-1}^2
 \right] \frac{t^k}{2} \, ,
\\
={}&
\sum_{k=0}^{2j} \binom{2j}{k} \binom{2j+1}{k} t^k \, .
\end{aligned}
\end{equation} 
By comparing Eqs.~\eqref{App.Eq0} and \eqref{App.Eq1}, we deduce that
\begin{equation}
\Delta_{k} =  (-1)^{2j-k} \binom{2j+1}{k} \, , \quad \text{for } k=0 , \, \dots , \, 2j \, .
\end{equation}
which is the result stated in Eq.~\eqref{Eig.P}.
\begin{table}[t!]
    \centering
    \begin{tabular}{c | c | l}
      \hspace{0.3cm} $N=2j$ \hspace{0.3cm} & \hspace{0.3cm} $L$ \hspace{0.3cm} & $F(L , \mu)$ for $\mu = 0 , \dots , L$ \hspace{0.5cm}
      \\[3pt]
      \hline      
       2  & $\phantom{\{ } \, 2$ &  $ \phantom{\left\{ \right.} \quad \frac{18}{35}, -\frac{24}{35}, \frac{6}{35}$ 
       \\[7pt]
       3  & $\left\{ \begin{array}{c}
            2  \\
            3 
       \end{array} \right.$ & $\left\{  \begin{array}{c}
            \frac{18}{35}, -\frac{24}{35}, \frac{6}{35} \\
            \frac{2}{3},-\frac{4}{7},-\frac{2}{7},\frac{4}{21} 
       \end{array}  \right.$
       \\[15pt]
       4  & $\left\{ \begin{array}{c}
            3  \\
            4 
       \end{array} \right.$ &
       $ \left\{ \begin{array}{c}
            \frac{100}{231},-\frac{50}{77},\frac{20}{77},-\frac{10}{231} \\
            \frac{250}{429},-\frac{90}{143},-\frac{20}{143},\frac{10}{39},-\frac{10}{143} 
       \end{array} \right. $
       \\[15pt]
       5  & $\left\{ 
       \begin{array}{c}
            3  \\
            4  \\
            5
       \end{array} 
       \right. $ &
       $ \left\{ \begin{array}{c}
            \frac{100}{231},-\frac{50}{77},\frac{20}{77},-\frac{10}{231} \\
            \frac{250}{429},-\frac{90}{143},-\frac{20}{143},\frac{10}{39},-\frac{10}{143}   
            \\
            \frac{2}{3},-\frac{80}{143},-\frac{40}{143},\frac{20}{429},\frac{30}{143},-\frac{12}{143}
       \end{array} 
       \right.$
    \end{tabular}
    \caption{Numerical values of $F(L, \mu)$ defined in Eq.~\eqref{F.fun} for several number of qubits and $N \geqslant L > N/2 $.}
    \label{tab.FLmu}
\end{table}
\section{Proof of Eq.~\texorpdfstring{\eqref{Global.min}}{Lg}}
\label{quadratic.minimum}
%
We begin by noting that $P_{\LB}(\rho)$, as given in Eq.~\eqref{general.PBL}, depends only on the components $\rho_{L0}$, and on $r$ which, however, is constant along the unitary orbit of $\rho$. These components can be expressed for a generic state within the unitary orbit, $\rho = U \Lambda U^{\dagger}$, with $\Lambda$ being a diagonal matrix in the $\ket{j,m}$-basis whose diagonal entries are the eigenvalues of $\rho$, as follows
\begin{equation}
    \rho_{L0} = \bm{\lambda} B \,\mathbf{t}_L^T  = \mathbf{v}_L \,\mathbf{b}'^T \, ,
\end{equation}
where $\mathbf{t}_L$ is the vector of eigenvalues of $T_{L0}$ with the entries $t_{L,m} = \bra{j,m} T_{L0} \ket{j,m}$, $B_{ij} = |U_{ij}|^2$ is a unistochastic matrix (hence also bistochastic) parametrized by a generalization of Eq.~\eqref{Par.b.bis}  in the $b_k$ variables, and the new variables $b_k= b'_k + \frac{1}{N+1} $ are defined to remove an irrelevant constant term. The vectors $\mathbf{v}_L$ have the components
\begin{equation}
    [\mathbf{v}_L]_k=\bm{\lambda} \frac{\partial B}{\partial b'_k} \,\mathbf{t}_L^T
\end{equation}
for $k=1 , \dots , N^2$, where the derivative of $B$ has all elements zero with the exception of
\begin{equation*}
\begin{aligned}
    & \bigg[\frac{\partial B}{\partial b'_k}\bigg]_{i,j}=\bigg[\frac{\partial B}{\partial b'_k}\bigg]_{N+1,N+1}=1\\
    &\bigg[\frac{\partial B}{\partial b'_k}\bigg]_{i,N+1}=\bigg[\frac{\partial B}{\partial b'_k}\bigg]_{N+1,j}=-1
\end{aligned}
\end{equation*}
where $i=\big\lfloor{\frac{k-1}{N}}+1\big\rfloor$ and $j \equiv k-1 \pmod{N}$. The function $P_{\LB}$ defined in Eq.~\eqref{general.PBL}, written as a quadratic function on the coefficients of $\mathbf{b}'$, then reads
\begin{equation}
\label{quad.problem.PLB}
P_{\LB} = f + 2\, \mathbf{q} \mathbf{b}'^T + {\mathbf{b}'} H \,\mathbf{b}'^T \, ,
\end{equation}
with
\begin{equation}
\label{hessian.B4}
\begin{aligned}
   & \mathbf{q} = \frac{1}{2} \sum_{k=1}^{N} g_k \mathbf{v}_k \, , \quad H = \sum_{k=1}^{N} h_k \mathbf{v}_k^T \mathbf{v}_k \, .
\end{aligned}
\end{equation}
We assume in Sec.~\ref{Sec.3.W2} that the eigenvalues $h_L$ of the matrix $H$ are non-negative. On the other hand, the vectors $ \mathbf{v}_1 , \dots , \mathbf{v}_L $ can be completed to form an orthogonal basis $\mathcal{V} = \{ \mathbf{v}_k \}_{k=1}^{N^2} $. We now define the dual basis $\tilde{\mathcal{V}} =\{ \tilde{\mathbf{v}}_k \}_{k=1}^{N^2}$ of $\mathcal{V}$ as the set of vectors satisfying
\begin{equation}
    \tilde{\mathbf{v}}_k \mathbf{v}_l^T = \delta_{kl} \, .
\end{equation}
We use this basis to define a new affine transformation $\mathbf{b}' = \mathbf{b}'' - \tilde{H} \mathbf{q}^T $ with
\begin{equation}
    \tilde{H} = \sum_{k=1}^{N} h_k^{-1} \tilde{\mathbf{v}}^T_k \tilde{\mathbf{v}}_k \, .
\end{equation}
Now, $P_{LB}$ written in terms of the new variables reads
\begin{equation}
\begin{aligned}
    P_{\LB} ={}& f
+{\mathbf{b}''} H \mathbf{b}''^{T} - \mathbf{q} \tilde{H} \mathbf{q}^T\\
={}&
f +{\mathbf{b}''} H \mathbf{b}''^{T} - \frac{1}{4} \sum_{k=1}^N \frac{g_k^2}{h_k} \, ,   
\end{aligned}
\end{equation}
where we have used that $H \tilde{H} \mathbf{q}^{T} = \mathbf{q}^{T}$. The global minimum of $ P_{\LB}$ is achieved for $\mathbf{b}''=\mathbf{0}$ because $H$ is a non-negative matrix, which proves the result~\eqref{Global.min}.
Note that the global minimum of $P_{LB}$ is strongly degenerate because we can move $\mathbf{b}''$ through the directions of the kernel $H$ without affecting the value of $P_{LB}$.
To illustrate this, let's look at the procedure described above for $N=2$. The vectors $\mathbf{v}_L$ are
\begin{equation}
\begin{aligned}    
    \mathbf{v}_1 = & \frac{1}{\sqrt{2}}\Big( 
    2(\la_0 - \la_2 ) , \la_0 - \la_2 , 2(\la_1 - \la_2 ) , \la_1 - \la_2   
    \Big) \, ,  
    \\
    \mathbf{v}_2 = & -\sqrt{\frac{3}{2}} \Big( 0, \la_0 -\la_2 , 0 , \la_1- \la_2 \Big) \, ,
\\
\mathbf{v}_3 = & \Big(\la_1 - \la_2 ,0, -\la_0 + \la_2 , 0 , \Big) \, , 
    \\
    \mathbf{v}_4 = & \Big( 0 , \la_1 - \la_2 ,0, -\la_0 + \la_2 ,  \Big) \, ,
\end{aligned}
\end{equation}
where we have completed the basis with two other vectors $\mathbf{v}_3$ and $\mathbf{v}_4$ which are orthogonal and satisfy the condition $H\mathbf{v}_3^T= H\mathbf{v}_4^T=0$. The corresponding vector $\mathbf{b}$ which satisfies the condition $\mathbf{b}''=0$ is
\begin{equation}
\begin{aligned}
& \mathbf{b}= \frac{\mathbf{1}}{3} + \left( 
\frac{35}{72 \big( \lambda _0^2+\lambda _1^2+2 \lambda _2^2-2 \left(\lambda _0+\lambda _1\right) \lambda _2 \big) y_2} \right) 
\\
& \times 
\Big( 5 (\la_0-\la_2), 4 (\la_2-\la_0), 5 (\la_1-\la_2), 4 (\la_2-\la_1) \Big)  
\\
& + z_3 \mathbf{v}_3 + z_4 \mathbf{v}_4
    \, ,
\end{aligned}
\end{equation}
where $\mathbf{1}$ is the vector with all the components equal to one, and $z_3,z_4$ are arbitrary real numbers.
\section{Properties of \texorpdfstring{$F(L,\mu)$}{Lg}}
\label{Appendix.Ineq.F}
%
Here we present some inequalities associated with the function $F(L,\mu)$ defined in Eq.~\eqref{F.fun}. 

\emph{Property 1.}
\begin{equation}\label{Ineq.1}
    F(2j,1) < 0
\end{equation} 

    \emph{Proof.} Writing $N=2j$, we have
    \begin{equation*}
        \begin{aligned}
            F(N,1) ={}& 2\sum_{\si=0}^{N} C_{N0N0}^{\si 0} C_{N1 N-1}^{\si 0}
            \\
            ={}& 2\sum_{\si=0}^{N} \left( C_{N0N0}^{\si 0} \right)^2 \left( \frac{\si(\si+1)}{2N(N+1)}-1 \right)
            \\
            <{}& 2\sum_{\si=0}^{N} \left( C_{N0N0}^{\si 0} \right)^2 \left( \frac{N(N+1)}{2N(N+1)}-1 \right)
            \\
            <{}& \,  0 \, , \quad \Box
        \end{aligned}
    \end{equation*}
   where we used an identity between the $C_{N0N0}^{\si0}$ and $C_{N1N-1}^{\si}$ given in Eq.~(7), p.~253 of Ref.~\cite{Var.Mos.Khe:88}.

\emph{Property 2.}
For all $\mu=0,\ldots,L$,
\begin{equation}
\label{Ineq.5}
    F(L,1) \leqslant F(L, \mu)  \, \text{ and } F(L,1)<0 \, .
\end{equation}
This inequality is a conjecture supported by the explicit evaluation of $F(L,\mu)$ for several $(L,\mu)$-values, as shown in Table~\ref{tab.FLmu} for $j\leq 5/2$. These inequalities become relevant for specifying the allowed domain of the $x_L$ parameters restricted by Eqs.~\eqref{restriction.ineq}-\eqref{restrict.hL}. If we assume that Eq.~\eqref{Ineq.5} is true, then the inequality that gives the best upper bound on  $x_L$ for $L > j$ comes from $\mu=1$
\begin{equation}
    x_L \leqslant \frac{F(2j,1)}{F(L,1)} x_{2j} \, ,
\end{equation}
where we have used that $F(2j,1) < 0$~\eqref{Ineq.1} and that $x_{2j}\geqslant 0$.
\section{Calculation of \texorpdfstring{$r_{\min} (\Sep_1)$}{Lg}}
\label{App.B}
%
We present here the calculation of $r_{\min}(\Sep_1)$ which, by convexity of $\Sep_1$, is equal to the radius $r_{\text{SAS}}$ of the largest inner ball within $\Sep_1$ centered on the maximally mixed state $\rho_0 = \left(N+1 \right)^{-1} \mathds{1}_{N+1}$ in the symmetric sector. In this Appendix, we closely follow the method proposed in Ref.~\cite{denis.martin:23}.
Before starting, it is convenient for our calculations to evaluate the norm of the $(2j+1)$-vector 
$\bm{\Delta}= (\Delta_{0}  , \dots , \Delta_{2j})$.
Using the explicit expression of the eigenvalues~\eqref{Eig.P}, we get
\begin{equation}
\label{Norm.eigenvalues}
    |\boldsymbol{\Delta}|^2 = \sum_{k=0}^{2j}\binom{2j+1}{k}^2 \, = \binom{4j+2}{2j+1}-1 \, ,
\end{equation}
where we used $\sum_{k=0}^{n}\binom{n}{k}^2=\binom{2n}{n}$ for the second equality. Then, we write the Hilbert-Schmidt distance \eqref{distance.rho.rho0} in terms of the Euclidean distance in the simplex between the spectra $\boldsymbol{\lambda}$ and $\boldsymbol{\lambda}_{0}$ of $\rho$ and $\rho_0$, respectively,
\begin{equation*}
r = \sqrt{\left(\sum_{i=0}^{2j}\lambda_{i}^{2}\right)-\frac{1}{2j+1}} = \lVert\boldsymbol{\lambda}-\boldsymbol{\lambda}_{0}\rVert \, .
\end{equation*}
By definition, $r_{\min}(\Sep_1)$ is the minimum distance between $\rho_0$ and the hyperplanes restricting the polytope $\Sep_1$. Mathematically, this translates in the following constrained minimization problem
\begin{equation}
\label{eq:Minimization_rin}
\min_{\boldsymbol{\lambda}} \; \lVert\boldsymbol{\lambda}-\boldsymbol{\lambda}_0 \rVert^2 \;\;\;
\text{ subject to } \left\{\begin{array}{l}
\sum_{i=0}^{2j}\lambda_{i}=1\\[8pt]
\boldsymbol{\lambda} \, \bm{\Delta}^T=0
\end{array}\right. \, .
\end{equation}
To solve this problem, we employ the method of Lagrange multipliers, introducing the Lagrangian 
\begin{equation*}
\mathcal{L} = \lVert\boldsymbol{\lambda}-\boldsymbol{\lambda}_0\rVert^2+\mu_{1}\boldsymbol{\lambda} \, \bm{\Delta}^T +\mu_{2}\bigg(1-\sum_{i=0}^{2j}\lambda_{i}\bigg).
\end{equation*}
Here, $\mu_{1}$ and $\mu_{2}$ represent Lagrange multipliers that will be determined. The stationary points, denoted as $\tilde{\boldsymbol{\lambda}}$, of the Lagrangian must satisfy the following condition
\begin{equation}
\label{eq:LagrangianDerivative}
\frac{\partial \mathcal{L}}{\partial\boldsymbol{\lambda}}\Big|_{\boldsymbol{\lambda}=\tilde{\boldsymbol{\lambda}}} = \boldsymbol{0} \quad \Leftrightarrow \quad 2\tilde{\boldsymbol{\lambda}}+\mu_{1}\boldsymbol{\Delta}-\mu_{2}\boldsymbol{1} = \boldsymbol{0}
\end{equation}
with the vector $\boldsymbol{1}$ which consists of elements all equal to 1 and has a length of $2j+1$. By summing across the components of \eqref{eq:LagrangianDerivative}, we get
\begin{equation}\label{mu2mu1}
\mu_{2} = \frac{\mu_{1}+2}{2j+1}.
\end{equation}
Next, by performing the scalar product of \eqref{eq:LagrangianDerivative} with $\boldsymbol{\Delta}$ and using \eqref{mu2mu1}, we obtain the following Lagrange multipliers:
\begin{equation*}
\mu_{1} = \frac{2}{(2j+1)|\boldsymbol{\Delta}|^2-1}, \quad \mu_{2} = \frac{2|\boldsymbol{\Delta}|^2}{(2j+1)|\boldsymbol{\Delta}|^2-1}.
\end{equation*}
Finally, upon substituting the above values for $\mu_1$ and $\mu_2$ into Eq.~\eqref{eq:LagrangianDerivative} and solving for the stationary point $\tilde{\boldsymbol{\lambda}}$, we arrive at the following expression:
\begin{equation}\label{rhostar}
\tilde{\boldsymbol{\lambda}}=\frac{|\boldsymbol{\Delta}|^2\mathbf{1}-\boldsymbol{\Delta}}{(2j+1)|\boldsymbol{\Delta}|^2-1}
\end{equation}
which leads us to
\begin{equation}\label{rminW1}
r_{\min}(\Sep_1) = \frac{1}{\sqrt{(2j+1)\Big[ (2j+1)|\boldsymbol{\Delta}|^2-1 \Big]}}
\end{equation} 
since $r_{\min}(\Sep_1) = r(\tilde{\rho})$ with $\tilde{\rho}$ any state with eigenspectrum \eqref{rhostar}. Remarkably, the radius \eqref{rminW1} is exactly the same as that deduced in Ref.~\cite{Boh.Gir.Bra:17} [see Eq.~\eqref{rmax.2017}], as can be verified by inserting Eq.~\eqref{Norm.eigenvalues} into \eqref{rminW1}.

We close this Appendix by noting that the same calculation generalizes directly to the general Stratonovich-Weyl kernel $w^{(s)}(\Omega)$ defined by 
\begin{equation}
\label{kernel.sgen}
w^{(s)}(\Omega) = \sqrt{\frac{4\pi}{2j+1}} \sum_{L=0}^{2j} \sum_{M=-L}^L \left( C_{j j L0}^{j j} \right)^{-s} Y_{LM}^{*} (\Omega) T_{LM} \, .
\end{equation}
where $s\in [-1,1] $~\cite{klimov2017gen}.
\end{appendix}
\bibliographystyle{apsrev4-2}
\bibliography{Refs_APP}
\end{document}